\documentclass[aps,pra,twocolumn,letterpaper,showpacs,superscriptaddress,floatfix,10pt]{revtex4-1}
\usepackage[pdftex]{graphicx}
\graphicspath{ {./images/} }
\usepackage{dcolumn}
\usepackage{bm}
\usepackage{bbm}
\usepackage[usenames, dvipsnames]{color}
\usepackage{comment}
\usepackage[colorlinks, linkcolor=blue]{hyperref}

\usepackage{layouts}

\usepackage[T1]{fontenc}

\usepackage{amsfonts}
\usepackage{amssymb}
\usepackage{amsmath}

\usepackage{amsthm}
\theoremstyle{remark}

\usepackage{multirow}

\usepackage[scr=boondoxo, scrscaled=1.05]{mathalfa}

\usepackage{array}
\newcolumntype{L}[1]{>{\raggedright\let\newline\\\arraybackslash\hspace{0pt}}m{#1}}
\newcolumntype{C}[1]{>{\centering\let\newline\\\arraybackslash\hspace{0pt}}m{#1}}
\newcolumntype{R}[1]{>{\raggedleft\let\newline\\\arraybackslash\hspace{0pt}}m{#1}}
\usepackage{tabularx}

\def\KeyWord#1{$\backslash$\IfColor{$\!\!$\textRed{#1}\textBlack}{#1}$\!\!$}

\usepackage{wasysym}

\newcommand{\reals}{\mathbb{R}}

\newcommand{\integers}{\mathbb{Z}}

\renewcommand{\d}{\mathrm{d}}

\newcommand{\Bn}{{\vec{n}}}
\newcommand{\Bm}{{\vec{m}}}
\newcommand{\Bk}{{\vec{k}}}
\newcommand{\Bl}{{\vec{\ell}}}

\newcommand{\BO}{{\vec{\Omega}}}
\newcommand{\Bt}{{\vec{\theta}}}

\def\bra#1{\langle#1|}
\def\ket#1{|#1\rangle}

\def\ketbra#1#2{|#1\rangle\langle#2|}
\def\qexp#1#2{\bra{#2}#1\ket{#2}}
\def\cexp#1{\langle#1\rangle}

\def\tr#1{\mathrm{Tr}\left[#1\right]}

\begin{document}
\title{Many-Body Localization with Quasiperiodic Driving}

\author{David M. Long}
\email{dmlong@bu.edu}
\affiliation{Department of Physics, Boston University, Boston, Massachusetts 02215, USA}

\author{Philip J. D. Crowley}
\affiliation{Department of Physics, Massachusetts Institute of Technology, Cambridge, Massachusetts 02139, USA}

\author{Anushya Chandran}
\affiliation{Department of Physics, Boston University, Boston, Massachusetts 02215, USA}

\date{\today}

\begin{abstract}
	Sufficient disorder is believed to localize static and periodically-driven interacting chains.
	With quasiperiodic driving by \(D\) incommensurate tones, the fate of this many-body localization (MBL) is unknown.
	We argue that randomly disordered MBL exists for \(D=2\), but not for \(D\geq 3\).
	Specifically, a putative two-tone driven MBL chain is neither destabilized by thermal avalanches seeded by rare thermal regions, nor by the proliferation of long-range many-body resonances.
	For \(D\geq 3\), however, sufficiently large thermal regions have continuous local spectra and slowly thermalize the entire chain.
	En route, we generalize the eigenstate thermalization hypothesis to the quasiperiodically-driven setting, and verify its predictions numerically.
	Two-tone driving enables new topological orders with edge signatures; our results suggest that localization protects these orders indefinitely.
\end{abstract}

\maketitle

\section{Introduction}
	\label{sec:intro}

	Strong periodic driving generates new phases of matter with no analog in static systems~\cite{Oka2019,Rudner2020,Rodriguez-Vega2021}. Examples include anomalous topological insulators with chiral edge modes~\cite{Titum2016,Po2016,Roy2017a,Nathan2017}, and discrete time crystals with sub-harmonic response to the drive~\cite{Wilczek2012,Khemani2016,Else2016}. Several optical and solid-state experiments have observed signatures of these dynamical phases~\cite{Zhang2017,Choi2017,Mi2021,Peng2016,Wintersperger2020}.

	Similarly, quasiperiodic driving by multiple incommensurate tones~\cite{Ho1983,Luck1988,Casati1989,Jauslin1991,Blekher1992,Jorba1992,Feudel1995,Bambusi2001,Gentile2003,Chu2004,Gommers2006,Chabe2008,Zhao2021} generates orders not accessible in either static \emph{or} periodically-driven systems~\cite{Mei2016,Martin2017,Nandy2017,Kolodrubetz2018,Peng2018,Lin2018,Petrides2018,Ray2019,Ozawa2019a,Else2019,Zhao2019,Crowley2020,Nathan2020b,Friedman2020,Long2021}, some of which have been experimentally observed~\cite{Lohse2018,Zilberberg2018,Lustig2018,Boyers2020,Dutt2020,Dumitrescu2021}. For instance, anomalous localized phases support energy currents between the drives at their edges~\cite{Kolodrubetz2018,Long2021,Nathan2020b}, and spin chains without any assumed symmetry exhibit coherent edge states~\cite{Friedman2020}. In both cases, the orders rely on localization in the bulk to forbid heating to a featureless infinite-temperature state~\cite{Anderson1958,Basko2006,Oganesyan2007,Pal2010,Serbyn2013,Huse2013,Huse2014,Ponte2015,Lazarides2015,Schreiber2015,Imbrie2016,Smith2016,Bordia2017,Leonard2020}.

	However, with interactions and quasiperiodic driving, it is not known if the bulk can remain localized indefinitely, and thus if these orders characterize genuine dynamical phases of  matter. Localization in quasiperiodically-driven systems is likely to be delicate, as even qubits can have ergodic dynamics and act as a local heat bath for nearby degrees of freedom~\cite{Jauslin1991,Blekher1992,Long2021,Nathan2020b}. Indeed, this is why there is no localization in classical spin chains~\cite{Oganesyan2009,Basko2011}.

	We provide analytical and numerical evidence that quasiperiodically-driven many-body localization (MBL) is a stable dynamical phase for smooth two-tone driving. Here, few-level systems generically have pure point spectra (\autoref{sec:synth}). Analogous arguments to those in static MBL then show that perturbations do not lead to the proliferation of long-range many-body resonances~\cite{Gopalakrishnan2015,Villalonga2020,Crowley2020b} (\autoref{sec:many_res}).

	\begin{figure}
		\centering
		\includegraphics[width=\linewidth]{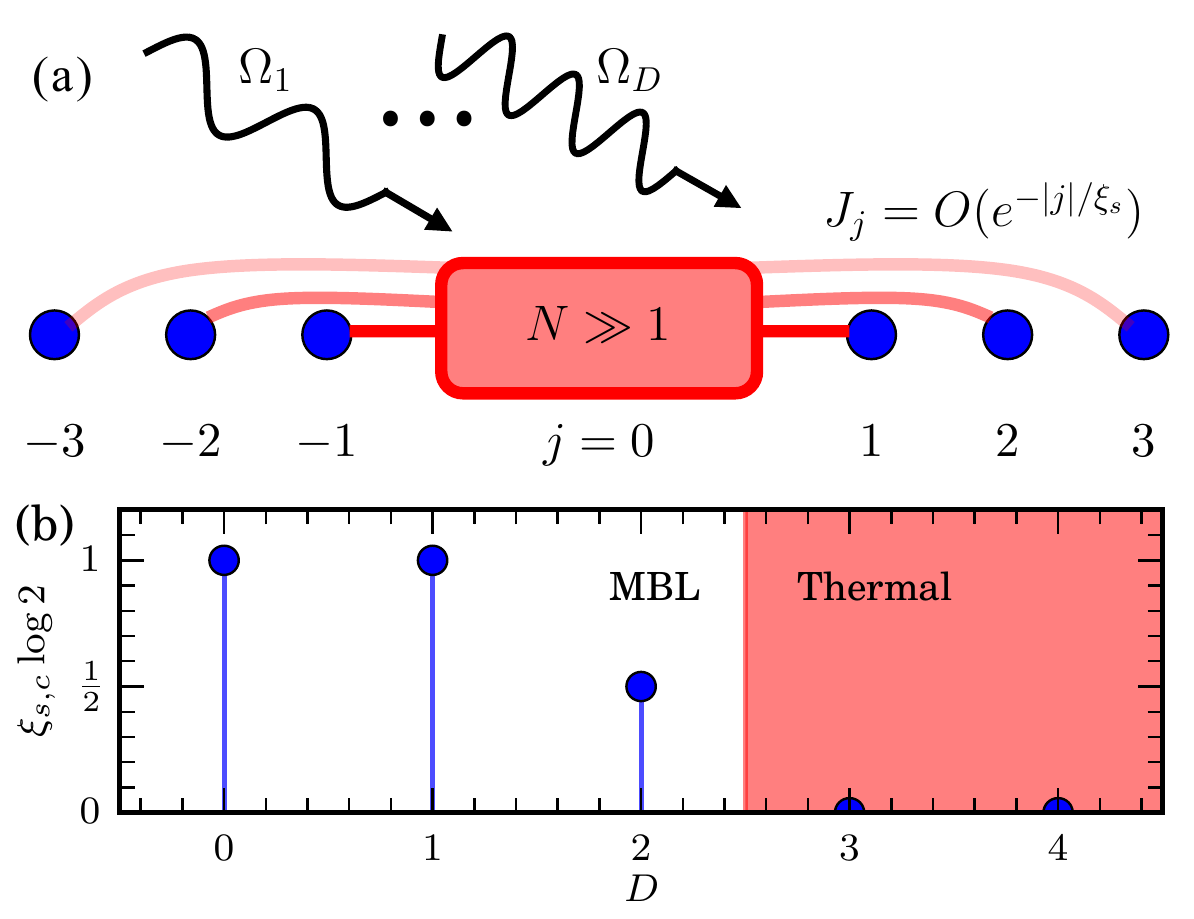}
		\caption{\label{fig:therm_incl}(\textbf{a})~\emph{Thermal inclusions.}--- The dominant mechanism of thermalization for a randomly disordered driven chain is the occurrence of a thermal region, say at site \(j=0\). The system is driven by \(D\) tones with frequencies \(\Omega_1, \ldots, \Omega_D\), and the \(N\)-level thermal region has exponentially decaying couplings \(J_j = O(e^{-|j|/\xi_s})\) to l-bits a distance \(j\) from the thermal region. (\textbf{b})~\emph{Critical localization length.}--- MBL is stable to the inclusion of a thermal region for \(D=0\) (static systems), \(D=1\) (periodically driven), \(D=2\), and not for any \(D \geq 3\). The critical localization length below which MBL is stable is reduced to \(\xi_{s,c} = (2\log 2)^{-1}\) for two-tone driving.}
	\end{figure}

	However, other potential instabilities remain -- in particular, for MBL by random disorder, a large thermal region with \(N\) levels may absorb nearby spins and initiate a thermal avalanche~\cite{deRoeck2017,Leonard2020} (\autoref{fig:therm_incl}). Here, the spectrum being pure point does not guarantee stability. Intuitively, the number of harmonics must grow slowly enough with \(N\) (\autoref{sec:space}). We show that the scaling with \(N\) allows for stable MBL when the localization length is less than a critical value,
	\begin{equation}
		\xi_{s,c} = (2 \log 2)^{-1}.
		\label{eqn:result}
	\end{equation}
	Notably, the critical localization length is reduced as compared to the static and periodically-driven cases~(\autoref{fig:therm_incl}).

	With three or more tones in the drive, sufficiently large thermal inclusions show continuous spectra~\cite{Long2021,Kolodrubetz2018}. Just as in classical systems, a putatively-MBL chain is not stable to such an inclusion. Thus, quasiperiodically-driven MBL with random disorder does not exist with three or more tones (\autoref{sec:more_tones}).

	Two of our intermediate results are of independent interest. We characterize l-bits with quasiperiodic driving (\autoref{sec:QPDMBL}) in terms of a frequency lattice which incorporates a synthetic dimension for each drive (\autoref{sec:back}). We also adapt the eigenstate thermalization hypothesis (ETH)~\cite{Jensen1985,Deutsch,Srednicki1994,Rigol2008,DAlessio2016} to quasiperiodically driven systems, and test its predictions numerically (\autoref{sec:ansatz}).

	In what follows, we focus on thermal inclusions in randomly disordered chains, before addressing the perturbative stability of MBL. The former is more constraining in its implications for MBL, and provides mathematical machinery with which to analyze the latter.

\section{Background -- Frequency Lattice}
	\label{sec:back}

	The \emph{frequency lattice} organizes the Fourier content of the long-time steady states of quasiperiodically-driven systems~\cite{Shirley1965,Sambe1973,Ho1983,Jauslin1991,Blekher1992,Verdeny2016}. It is well suited to discussions of formally infinite-time properties, such as localization. This section reviews the frequency lattice construction.

	We consider one-dimensional quantum systems with smooth quasiperiodic time dependence consisting of \(D\) incommensurate tones. Such a Hamiltonian may be parameterized in terms of \(D\) phase variables \(\theta_j(t) = \Omega_j t\), where \(\Omega_j\) is the angular frequency of the \(j\)th drive. For convenience, we assemble the phases into a vector
	\begin{equation}
		\Bt_t = \sum_{j=1}^D \theta_j(t) \hat{e}_j.
	\end{equation}
	The time-dependent Hamiltonian may then be written as
	\begin{equation}
		H(t) = H(\Bt_t), \quad\text{where}\quad H(\Bt + 2\pi \hat{e}_j) = H(\Bt)
	\end{equation}
	is periodic in each phase variable, with period \(2\pi\). Incommensurability of the frequencies is stated mathematically as
	\begin{equation}
		\Bn \cdot \BO = 0 \quad\Longleftrightarrow\quad \Bn = 0,
	\end{equation}
	where \(\Bn \in \integers^D\) is a vector of integers. (For \(D=2\), this is equivalent to \(\Omega_1/\Omega_2\) being irrational.) The drive is not periodic, but is instead, in a sense that can be made precise, almost periodic.

	In analogy to the stationary state solutions of the Schr\"odinger equation with a static Hamiltonian, the steady states of a quasiperiodically driven system are the \emph{quasienergy states}~\cite{Blekher1992,Floquet1883,Ho1983}
	\begin{equation}
		\ket{\psi_\alpha(t)} = e^{-i\epsilon_\alpha t}\ket{\phi_\alpha(\Bt_t)},
	\end{equation}
	where \(\ket{\psi_\alpha(t)}\) is a solution to the Schr\"odinger equation \(i\partial_t \ket{\psi_\alpha(t)} = H(t)\ket{\psi_\alpha(t)}\), \(\alpha\) indexes a basis of the Hilbert space, \(\epsilon_\alpha\) is the \emph{quasienergy} and the quasienergy state \(\ket{\phi_\alpha(\Bt_t)}\) is smooth on the torus. The states
	\begin{equation}
		\ket{\phi_\alpha(\Bt)} = \sum_{\Bn \in \integers^D} \ket{\phi_{\alpha \Bn}} e^{-i\Bn\cdot \Bt}
	\end{equation}
	may be calculated after a Fourier transform from the eigenvalue equation
	\begin{equation}
		\sum_{\Bm \in \integers^D} K_{\Bn \Bm} \ket{\phi_{\alpha \Bm}} = \epsilon_\alpha \ket{\phi_{\alpha \Bn}},
	\end{equation}
	where
	\begin{equation}
		K_{\Bn \Bm} = H_{\Bn - \Bm} - \BO \cdot \Bn \delta_{\Bn \Bm},
	\end{equation}
	and \(H_{\Bn}\) are the Fourier components of \(H(\Bt) = \sum_{\Bn} H_{\Bn} e^{-i\Bn\cdot\Bt}\). The quasienergy states being smooth on the torus is equivalent to the Fourier components \(\ket{\phi_{\alpha \Bn}}\) being localized in \(\Bn\). If the eigenstates of \(K_{\Bn\Bm}\) are delocalized, the quasienergy states are not well-defined.

	The operator \(K_{\Bn \Bm}\) is a static lattice Hamiltonian in an extended \emph{frequency lattice}. It has translationally-invariant hopping matrices given by \(H_{\Bn - \Bm}\), and an on-site linear potential \(-\BO \cdot \Bn\) which breaks translational symmetry. This linear potential would arise in real-space from a uniform electric field given by \(\BO\), so we sometimes refer to \(\BO\) in this context as an electric field.

	\begin{figure}
		\centering
		\includegraphics[width=\linewidth]{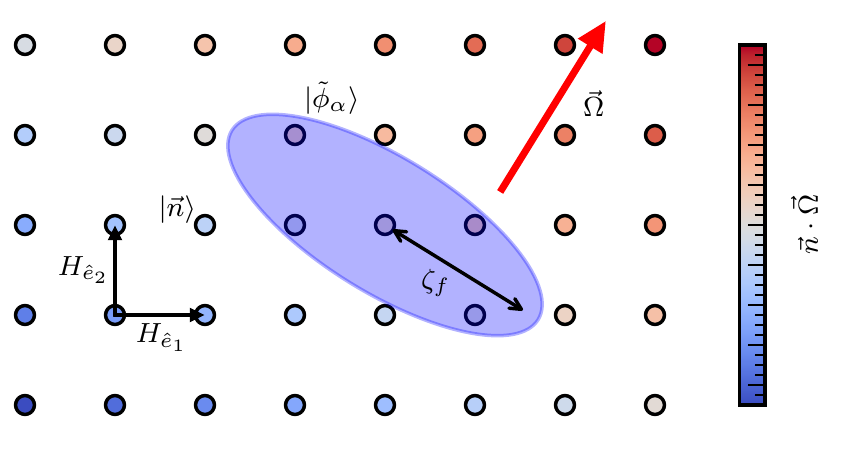}
		\caption{\label{fig:frq_lat}\emph{The frequency lattice.}--- The steady states of a system driven by \(D\) incommensurate tones are the eigenstates of a static lattice problem in an extended frequency lattice. This lattice has additional synthetic dimensions, with sites labeled by \(\Bn \in \integers^D\) (each site shown has all of the degrees of freedom of the spatial Hilbert space). The hopping matrices \(H_{\Bn - \Bm}\) in the frequency lattice are given by Fourier components of the driven Hamiltonian. The on-site linear potential is \(-\Bn\cdot\BO\), as might arise from a uniform electric field \(\BO\). The quasienergy states \(\ket{\tilde{\phi}_\alpha}\) are localized with localization length \(\zeta_f\). The degree of localization parallel to \(\BO\) is greater than that perpendicular to \(\BO\).}
	\end{figure}

	The frequency lattice has additional \emph{synthetic dimensions} corresponding to each of the periodic drives (\autoref{fig:frq_lat}). We make this explicit by appending states \(\ket{\Bn}\) to the Hilbert space and defining~\cite{Ho1983,Jauslin1991,Blekher1992,Verdeny2016}
	\begin{equation}
		\tilde{K} = \sum_{\Bn, \Bm \in \integers^D} K_{\Bn \Bm} \ketbra{\Bn}{\Bm},
	\end{equation}
	and similarly \(\ket{\tilde{\phi}_\alpha} = \sum_{\Bn} \ket{\phi_{\alpha\Bn}}\ket{\Bn}\). Explicitly, the extended Hilbert space is
	\begin{equation}
		 \mathcal{K} = \mathcal{H}\otimes \ell^2(\integers^D),
	\end{equation}
	where \(\mathcal{H}\) is the Hilbert space in the temporal domain, and \(\ell^2(\integers^D)\) denotes the space of square-summable complex valued functions on the square lattice \(\integers^D\).

	We will decorate states in, and operators on, \(\mathcal{K}\) with a tilde, to make a clear distinction between those objects that have the extra factor \(\ell^2(\integers^D)\), and those that do not.

	Extending the Hilbert space introduces a new gauge freedom related to the position of the origin in the synthetic dimensions. Translations in the synthetic dimensions do not produce observable effects on real-time dynamics, as may be seen explicitly from the quasienergy states. A translation of a quasienergy state \(\ket{\tilde{\phi}_\alpha}\) by a lattice vector \(\Bm\),
	\begin{equation}
		\ket{\tilde{\phi}_\alpha^\Bm} = \sum_{\Bn} \ket{\phi_{\alpha\Bn}}\ket{\Bn + \Bm},
		\label{eqn:qenst_trans}
	\end{equation}
	is another quasienergy state of \(\tilde{K}\), with quasienergy \(\epsilon_\alpha - \Bm \cdot \BO\). The actual solution to the Schr\"odinger equation, however, does not change:
	\begin{equation}
		\ket{\psi_\alpha^\Bm(t)} = e^{-i(\epsilon_\alpha - \Bm \cdot \BO)t}e^{-i \Bm \cdot \BO t}\ket{\phi_\alpha(\Bt)} = \ket{\psi_\alpha(t)}.
	\end{equation}

	An operator \(O(\Bt)\) on \(\mathcal{H}\) corresponds to an operator on \(\mathcal{K}\) defined by
	\begin{equation}
		\tilde{O} = \sum_{\Bn,\Bm} O_{\Bn-\Bm}\ketbra{\Bn}{\Bm},
		\label{eqn:obs_freq}
	\end{equation}
	which is constructed so that \(O(\Bt)\ket{\phi(\Bt)} \leftrightarrow \tilde{O} \ket{\tilde{\phi}}\). We see that physical operators are naturally translationally invariant (gauge invariant) in the frequency lattice.

	When \(\mathcal{H}\) is a many-body Hilbert space for a spatially extended system, the character of the spatial dimensions is different from the synthetic frequency lattice dimensions. If we consider a finite subsystem of the frequency lattice for a spin-\(\tfrac{1}{2}\) chain with \(L\) spins and \(M\) synthetic sites, the Hilbert space dimension supported on this subsystem is \(2^L M\). The synthetic part of the problem is thus analogous to a single-particle system, even in the many-body setting.

	Furthermore, the structure of tensor products in the frequency lattice is more complicated than in the temporal domain. The origin of this complication is that there is only one factor of \(\ell^2(\integers^D)\) in the frequency lattice Hilbert space, even in a tensor product system. Explicitly, if \(\mathcal{H} = \mathcal{H}_1 \otimes \mathcal{H}_2\), then
	\begin{equation}
		\mathcal{K} = \mathcal{H}_1 \otimes \mathcal{H}_2 \otimes \ell^2(\integers^D) \neq \mathcal{K}_1 \otimes \mathcal{K}_2,
	\end{equation}
	where \(\mathcal{K}_j = \mathcal{H}_j \otimes \ell^2(\integers^D)\). As a consequence, given states \(\ket{\phi_j(\Bt)} \in \mathcal{H}_j\) and corresponding frequency lattice states \(\ket{\tilde{\phi}_j} \in \mathcal{K}_j\), the frequency lattice state corresponding to \(\ket{\phi_1(\Bt)}\otimes \ket{\phi_2(\Bt)}\) is obtained as a convolution, for which we use the symbol \(*\),
	\begin{equation}
		\ket{\tilde{\phi}_1 \tilde{\phi}_2} = \ket{\tilde{\phi}_1} * \ket{\tilde{\phi}_2}
		= \sum_{\Bn}\left(\sum_{\Bm} \ket{\phi_{1,\Bn-\Bm}}\ket{\phi_{2,\Bm}} \right)\ket{\Bn},
	\end{equation}
	and not as a tensor product of the states \(\ket{\tilde{\phi}_j}\).
	
	Such tensor convolutions are somewhat more elegantly stated for operators. An operator \(O_j(\Bt)\) on \(\mathcal{H}_j\) corresponds to an operator \(\tilde{O}_j\) on \(\mathcal{K}_j\) defined as in Eq.~\eqref{eqn:obs_freq}. The frequency lattice operator for the tensor product \(O_1(\Bt)\otimes O_2(\Bt)\) is
	\begin{equation}
		\widetilde{O_1 O_2} = (\tilde{O}_1\otimes\mathbbm{1}_2)(\mathbbm{1}_1\otimes \tilde{O}_2) =  \tilde{O}_1 \tilde{O}_2 = \tilde{O}_2\tilde{O}_1 ,
	\end{equation}
	where in the last two expressions we use the convention that \(\tilde{O}_1\) acting in \(\mathcal{K}\) is regarded as acting as the identity on the space \(\mathcal{H}_2\), and similarly for \(\tilde{O}_2\) acting on \(\mathcal{H}_1\).

\section{Quasiperiodically-driven Many-Body Localization}
	\label{sec:QPDMBL}

	We present a definition of MBL in a quasiperiodically driven setting that recovers much of the phenomenology present in static systems. In static systems, MBL may be characterized by a complete set of quasilocal integrals of motion, \emph{l-bits} \(\tau^z_j\), for which
	\begin{equation}
		\qexp{\tau^z_j}{\psi(t)} = \mathrm{const.}
	\end{equation}
	for any initial state \(\ket{\psi(0)}\). This property results in the many striking features of MBL: memory of the initial state, pure point spectra of local observables, and so on~\cite{Serbyn2013,Huse2014,Imbrie2016}.

	Similarly, we define a complete set of l-bits \(\tau^z_j(\Bt)\) with explicit \(\Bt\) dependence. The l-bits commute with the time evolution operator, so that
	\begin{equation}
		\qexp{\tau^z_j(\Bt_t)}{\psi(t)} = \mathrm{const.}
		\label{eqn:lbits_cons}
	\end{equation}
	for any initial state \(\ket{\psi(0)}\).

	A quasiperiodically driven system is many-body localized if there is a complete set of l-bits that are (quasi)local in both the frequency and spatial lattices. That is, a set of frequency lattice operators
	\begin{equation}
		\tilde{\tau}^z_j = \sum_{\Bn,\Bm \in \integers^D} \tau^z_{j, \Bn-\Bm} \ketbra{\Bn}{\Bm}
	\end{equation}
	such that \([\tilde{\tau}^z_j, \tilde{K}]=0\), \([\tilde{\tau}^z_j, \tilde{\tau}^z_k]=0\), and with \(\tilde{\tau}^z_j\) having localization center \(j\). More precisely, decomposing \(\tau^z_{j,\Bn}\) into terms \(\tau^z_{j,\Bn,r}\) supported within a spatial range \(r\) of \(j\):
	\begin{equation}
		\tau^z_{j,\Bn} = \sum_r \tau_{j,\Bn,r}^z
		\quad\text{where}\quad
		\|\tau^z_{j, \Bn, r}\| = O(e^{-|\Bn|/\zeta_f - r/\zeta_s}).
	\end{equation}
	Here, we have introduced a frequency localization length \(\zeta_f\), and a spatial localization length \(\zeta_s\).

	The complete set of l-bits split the Hilbert space into \(2^L\) sectors (for a spin-\(\tfrac{1}{2}\) chain of length \(L\)). Each sector contains only one physically inequivalent quasienergy state, and may be labeled by its eigenvalues under each \(\tilde{\tau}^z_j\). Furthermore, we require these quasienergy states to be localized in the synthetic dimensions -- that is, that they have smooth quasiperiodic time dependence in the temporal domain~\footnote{Our definition requires the quasienergy states \(\ket{\phi_\alpha(\Bt)}\) to be smooth on the torus. From Eq.~\eqref{eqn:lbit_temporal} we can see that the requirement that \(\tau^z_j(\Bt)\) be smooth (that is, that \(\tilde{\tau}_j^z\) be a quasilocal operator) implies that the projector \(\ketbra{\phi_\alpha(\Bt)}{\phi_\alpha(\Bt)}\) must be smooth. Even so, the requirement that \(\ket{\phi_\alpha(\Bt)}\) be smooth is an independent assumption which excludes the case of \(\ket{\phi_\alpha(\Bt)}\) not admitting a globally smooth gauge -- for instance, because it has a non-trivial Chern number.}.

	Explicitly, if we label the quasienergy state \(\ket{\tilde{\phi}_\alpha}\) translated by the frequency lattice vector \(\Bn\) as \(\ket{\tilde{\phi}_\alpha^\Bn}\), then the frequency lattice l-bits may be written as
	\begin{equation}
		\tilde{\tau}^z_j = \sum_{\Bn, \alpha} \tau^z_{j\alpha} \ketbra{\tilde{\phi}_\alpha^\Bn}{\tilde{\phi}_\alpha^\Bn},
		\label{eqn:tau_quasist}
	\end{equation}
	where \(\tau^z_{j\alpha}\) is an \(\Bn\) independent eigenvalue (recall that any physical operator must be translationally invariant in the synthetic dimensions). Eq.~\eqref{eqn:tau_quasist} also makes clear that the frequency lattice localization length of the l-bits, \(\zeta_f\), is also that of the quasienergy states (\autoref{fig:frq_lat}).

	In later sections, we only use the frequency lattice operators \(\tilde{\tau}^z_j\). The corresponding temporal operators are conserved quantities with explicit time dependence, as we show below.

	In the temporal domain, \(\tilde{\tau}^z_j\) corresponds to a smooth, quasilocal, quasiperiodic operator
	\begin{equation}
		 \tau^z_j(t) = \tau^z_j(\Bt_t) = \sum_{\Bn} \tau^z_{j,\Bn}e^{-i\Bn\cdot\Bt_t}
	\end{equation}
	such that \(\tau^z_j(\Bt)\ket{\phi_\alpha(\Bt)} = \tau^z_{j \alpha}\ket{\phi_\alpha(\Bt)}\). That is
	\begin{equation}
		\tau^z_j(\Bt) = \sum_{\alpha} \tau^z_{j \alpha}\ketbra{\phi_\alpha(\Bt)}{\phi_\alpha(\Bt)}
		\label{eqn:lbit_temporal}
	\end{equation}
	is diagonal in the quasienergy state basis, even in the temporal domain.

	The temporal domain operators do not necessarily commute with the instantaneous Hamiltonian, \([\tau^z_j(\Bt), H(\Bt)] \neq 0\). Rather, the Heisenberg operators
	\begin{equation}
		\tau_j^{z,H}(t) = U(t)^\dagger\tau^z_j(\Bt_t)U(t)
	\end{equation}
	(where \(U(t) = U(t,0)\) is the unitary evolution operator) are constant in time
	\begin{equation}
		\d_t \tau_j^{z,H}(t) = 0
	\end{equation}
	so that the l-bits are conserved quantities with explicit time dependence. Taking an expectation value in \(\ket{\psi(0)}\) yields Eq.~\eqref{eqn:lbits_cons}.

	Unlike in static MBL, the Hamiltonian \(H(\Bt)\) cannot be expressed as a sum of products of the l-bits~\cite{Serbyn2013,Huse2014}.
	Instead, the quasienergy operator in the frequency lattice has the analogous property that there exists a quasilocal unitary \(\tilde{W}\) in the frequency lattice so that
	\begin{multline}
		\tilde{W}\tilde{K}\tilde{W}^\dagger = -\sum_{\Bn} \BO \cdot \Bn \ketbra{\Bn}{\Bn} \\
		+ \sum_j h_j \tilde{\sigma}^z_j + \sum_{j,j'} h_{jj'} \tilde{\sigma}^z_j \tilde{\sigma}^z_{j'} + \cdots.
		\label{eqn:Ktau}
	\end{multline}
	That is, a quasilocal rotation allows \(\tilde{K}\) to be expressed as a sum of products of Pauli \(\tilde{\sigma}^z\) operators, up to a term that breaks the translational invariance.

	The definition of MBL implies that all local observables \(O\) have pure point power spectra~\cite{Jauslin1991,Blekher1992}, as is the case in static MBL.

	Ref.~\cite{Else2019} also proposes a definition of quasiperiodically-driven MBL. We show in Appendix~\ref{app:equiv} that the two definitions are equivalent.

\section{Thermal Region Ansatz}
	\label{sec:ansatz}

	In this section, we present an ansatz which characterizes matrix elements of \emph{thermalizing quasiperiodically-driven systems}, in the style of the eigenstate-thermalization hypothesis (ETH)~\cite{Jensen1985,Deutsch,Srednicki1994,Rigol2008,DAlessio2016}. This ansatz characterizes low-disorder regions in a quasiperiodically-driven putatively-MBL chain.

	Our ansatz is a statistical description of finite quasiperiodically-driven quantum systems with pure point spectra. In the thermodynamic limit, the spectrum becomes continuous. However, it is also possible to have a continuous spectrum in a finite quasiperiodically-driven system for \(D \geq 3\) (\autoref{sec:more_tones}). To develop an ETH ansatz here, the spectrum should be made discrete with commensurate approximations (\autoref{sec:disc}).

	Consider an \(N\)-dimensional Hilbert space with a quasiperiodic Hamiltonian \(H_B(\Bt_t)\) (the ``bath Hamiltonian''). Assume that there exists a complete set of smooth quasienergy states \(\ket{\psi_\alpha(t)} = e^{-i \epsilon_\alpha t}\ket{\phi_\alpha(\Bt_t)}\) -- that is, that the eigenstates of the corresponding quasienergy operator are localized in the synthetic dimensions, with localization length \(\zeta_f\).

	The ansatz concerns matrix elements of generic local operators \(V(\Bt_t)\) between quasienergy states,
	\begin{align}
		V_{\alpha \beta}(t) &= \bra{\psi_\alpha(t)} V(\Bt_t) \ket{\psi_\beta(t)} \nonumber \\
		&= e^{-i\omega_{\beta \alpha} t} \bra{\phi_\alpha(\Bt_t)} V(\Bt_t) \ket{\phi_\beta(\Bt_t)},
	\end{align}
	where \(\omega_{\beta \alpha} = \epsilon_\beta - \epsilon_\alpha\), and we choose particular representative quasienergy states \(\ket{\phi_\alpha(\Bt_t)}\) to fix \(\epsilon_\alpha\).

	The frequency lattice operator corresponding to \(V\) is \(\tilde{V} = \sum_{\Bn,\Bm} V_{\Bn-\Bm}\otimes\ketbra{\Bn}{\Bm}\), and the quasienergy states are denoted \(\ket{\tilde{\phi}_\alpha^\Bn}\)~\eqref{eqn:qenst_trans}. Then an arbitrary matrix element of \(\tilde{V}\) in the quasienergy state basis has the form
	\begin{align}
		\tilde{V}_{\alpha \beta}^{\Bn \Bm} &= \bra{\tilde{\phi}_\alpha^\Bn} \tilde{V} \ket{\tilde{\phi}_\beta^\Bm} \nonumber \\
		&= \sum_{\vec{j},\vec{k}} \bra{\phi_{\alpha\vec{j}}} V_{\vec{j}-\vec{k}+\Bn-\Bm} \ket{\phi_{\beta\vec{k}}},
	\end{align}
	which is the coefficient of \(\delta(\omega - \Delta^{\Bn \Bm}_{\alpha \beta})\) in the Fourier transform of \(V_{\alpha \beta}(t)\), and \(\Delta^{\Bn \Bm}_{\alpha \beta} = \omega_{\beta\alpha} + (\Bn - \Bm)\cdot \BO\) is the quasienergy difference between \(\ket{\tilde{\phi}_\beta^\Bm}\) and \(\ket{\tilde{\phi}_\alpha^\Bn}\). As \(\tilde{V}\) is translationally invariant, the matrix element \(\tilde{V}_{\alpha \beta}^{\Bn \Bm} = \tilde{V}_{\alpha \beta}^{\Bn-\Bm}\) only depends on the separation between \(\Bn\) and \(\Bm\), which we call \(\Bl = \Bn - \Bm\). Subsequently, we only keep the \(\Bl\) dependence in our notation.

	We first state the ansatz, and then define and motivate each of the terms appearing in the equation. The ansatz is
	\begin{equation}
		\tilde{V}_{\alpha \beta}^{\Bl} = \bar{V}_{\Bl} \delta_{\alpha \beta} + \frac{f_{\Bl}(\Delta^{\Bl}_{\alpha \beta})}{\sqrt{N \xi_f^{D-1}}} R_{\alpha \beta,\Bl}.
		\label{eqn:RMT_ans}
	\end{equation}

	Consider the first term. Equation~\eqref{eqn:RMT_ans} must recover the infinite-temperature expectation value of \(V(\Bt)\) in a quasienergy state, as it models a thermal system. (As energy is not conserved in a quasiperiodically driven system, thermal expectation values should be taken at infinite temperature.) We define \(\bar{V}(\Bt)\) to be this expectation value,
	\begin{equation}
		\bar{V}(\Bt) = \frac{1}{N}\tr{V(\Bt)} = \sum_{\Bl} \bar{V}_{\Bl} e^{-i\Bl\cdot \Bt}.
	\end{equation}
	The Fourier components \(\bar{V}_{\Bl}\) appear in the first term of Eq.~\eqref{eqn:RMT_ans}. Fluctuations to the expectation value are given by the second term in Eq.~\eqref{eqn:RMT_ans}, but these vanish as the number of levels \(N \to \infty\).
	
	The second term is motivated by the intuition that the components \(\ket{\phi_{\alpha\Bn}}\) appear as independent random vectors~\cite{DAlessio2016}, with an assumed exponentially decaying norm with \(|\Bn|\)~(\autoref{fig:frq_lat}).

	The factors \(R_{\alpha\beta,\Bl}\) are independent (usually complex) random variables with mean zero and unit variance, and model the apparently random nature of the quasienergy states. We will not need to assume any particular distribution for these variables, or even that they are identically distributed for different \(\Bl\). However, if \(V(\Bt)\) is Hermitian, then there is a constraint \(R_{\alpha \beta,\Bl} = R^*_{\beta \alpha,-\Bl}\), where \(z^*\) is the complex conjugate of \(z\).

	The spectral functions \(f_{\Bl}(\omega)\) appearing in the second term encode the dependence of the off-diagonal matrix elements on the quasienergy difference \(\omega\). The spectral functions also carry an explicit dependence on the frequency lattice separation \(\Bl\). The former is usual for an ETH ansatz -- matrix elements typically depend on energy differences of eigenstates. The latter dependence on \(\Bl\) has no analog in the usual ETH for static or periodically-driven systems -- it encodes the localization of the quasienergy states (and hence the matrix elements) perpendicular to \(\BO\) in the frequency lattice. Displacements \(\Bl\) parallel to \(\BO\) affect the quasienergy difference \(\omega = \Delta^{\Bl}_{\alpha \beta}\), but those perpendicular to \(\BO\) do not. As \(\omega\) is insensitive to this displacement, the additional dependence of \(f_{\Bl}(\omega)\) on \(\Bl\) is required to correctly describe the localization perpendicular to \(\BO\). Namely, for large \(|\Bl|\), we demand that
	\begin{equation}
		|f_{\Bl}(\omega)| = O(e^{-|\Bl|/\xi_f}),
	\end{equation}
	where \(\xi_f\) is a frequency lattice localization length. If \(V_{\Bn} = O(e^{-|\Bn|/\zeta_V})\), then \(\xi_f = O(\max\{\zeta_f, \zeta_V\})\). When the localization length of the quasienergy states is large, \(\xi_f = O(\zeta_f)\).

	The localization of \(f_{\Bl}\) in the direction parallel to the electric field \(\BO\) in the frequency lattice is much stronger than in the \(D-1\) perpendicular directions. This is due to Stark localization by the linear potential \(\Bn \cdot \BO\), which causes a super-exponential localization like
	\begin{equation}
		\log|f_\Bl(\omega)| \sim -\omega_\Bl \log\omega_\Bl,
		\label{eqn:nlogn}
	\end{equation}
	where \(\omega_\Bl = \Bl\cdot\hat{\Omega}\) is much larger than a localization length parallel to the electric field, \(\omega_\Bl \gg \xi_\parallel\)~\cite{Emin1987}.

	The localization length \(\xi_\parallel\) controls the preasymptotic exponential decay of \(|f_\Bl(\omega)|\), and depends only weakly on \(N\). In a driven many-body system, \(\xi_\parallel\) is a function of \(W/|\BO|\), where \(W\) is the bandwidth of the static part of the Hamiltonian. For a generic spin system this varies as \(W = O(\sqrt{L}) = O(\sqrt{\log_2 N})\), which results in a very weak growth with \(N\). States at a distance \(\omega_{\Bl} \gg \xi_\parallel\) are far detuned, resulting in super-exponential localization.

	The localization length \(\xi_f\) also appears in the denominator of the second term in Eq.~\eqref{eqn:RMT_ans}, which may be interpreted as the square root of an effective Hilbert space dimension
	\begin{equation}
		N_{\mathrm{eff}} = N \xi_f^{D-1}.
	\end{equation}
	For a given \(\ket{\tilde{\phi}_\alpha^\Bn}\), \(N_{\mathrm{eff}}\) is roughly the number of other states with which \(\ket{\tilde{\phi}_\alpha^\Bn}\) has a significant matrix element.
	More precisely, the volume factor of \(\xi_f^{D-1}\) in \(N_{\mathrm{eff}}\) ensures that
	\begin{equation}
		 \sum_{\alpha} \qexp{\tilde{V}^\dagger \tilde{V}}{\tilde{\phi}_\alpha} = \int \frac{\d^D \theta}{(2\pi)^D} \tr{V^\dagger V(\Bt)} = O(N).
	\end{equation}
	The exponent is \(D-1\), and not \(D\), because the strong localization in the \(\BO\) direction means that the relevant volume is (asymptotically for large \(N\)) just that perpendicular to \(\BO\).

	The predictions of our ansatz~\eqref{eqn:RMT_ans} can be checked in numerical simulations of thermalizing quasiperiodically-driven systems. In Appendix~\ref{app:numerical_ans} we check several statistics of the off-diagonal matrix elements of an operator between quasienergy states for \(D=2\), and find that they are consistent with~\eqref{eqn:RMT_ans}.

	As a final comment, there may be \(D-1\) distinct localization lengths in the plane perpendicular to \(\BO\) in the frequency lattice. Eq.~\eqref{eqn:RMT_ans} is modified accordingly; specifically, \(\xi_f^{D-1}\) is replaced with the product of principal localization lengths, \(\prod_{j=1}^{D-1} \xi_{f,j}\). More generally, this denominator is determined by the requirement of normalization. In later sections, we neglect such refinements and use Eq.~\eqref{eqn:RMT_ans} as stated, as our primary focus is \(D=2\), where there is a unique localization length \(\xi_f\) perpendicular to \(\BO\).

\section{Spatial Localization Assuming Synthetic Localization}
	\label{sec:space}

	In this section, we show that quasiperiodically-driven MBL is self-consistently stable to the inclusion of a thermal region, provided the frequency lattice localization length grows at most as a power law with the Hilbert space dimension of the thermal region, \(\xi_f = O(N^\nu)\).

	Intuitively, in the ETH ansatz~\eqref{eqn:RMT_ans} the effective density of states grows as
	\begin{equation}
		\rho_{\mathrm{eff}} = O(N_{\mathrm{eff}}) = O(N^{1 + \nu(D-1)}).
	\end{equation}
	For MBL to be self-consistently stable, the product of this density of states and a typical matrix element of a perturbation must be much less than unity. Testing when this is true, as in Ref.~\cite{deRoeck2017}, leads to the conclusion that MBL may be stable for spatial localization lengths obeying
	\begin{equation}
		\xi_s < \xi_{s,c} = ([1 + \nu(D-1)]\log 2)^{-1}.
		\label{eqn:crit_loc_summary}
	\end{equation}
	Eq.~\eqref{eqn:crit_loc_summary} is our main result of this section.

	A technical proof of Eq.~\eqref{eqn:crit_loc_summary} is more involved, as the density of states in the frequency lattice is formally infinite at all energies, and the matrix elements \(\tilde{V}^{\Bl}_{\alpha \beta}\) do not have a single scale. To characterize precisely how the infinite density of states is defeated by exponential localization in the matrix elements, we use the \emph{fidelity susceptibility} in the frequency lattice. The typical value of this quantity is
	\begin{equation}
		\chi_{\star} = \left(\lim_{\Delta \to 0}\left[\frac{1}{2\Delta}\sum_{|\Delta_{\beta\alpha}^{\Bl}|<\Delta} |\tilde{V}^{\Bl}_{\beta\alpha}|\right]\right)^2,
		\label{eqn:chi_star_intro}
	\end{equation}
	where the sum is over states in a narrow quasienergy window \(\Delta\), and the square brackets indicate an ensemble average, which we discuss further below. This quantity is well-defined in the frequency lattice.
	
	In a static system \(\sqrt{\chi_\star} = \rho [|V|]\) reduces to the familiar product of the density of states \(\rho\) and the average (absolute value of the) off-diagonal matrix element~\cite{Crowley2021}.

	We note that Eq.~\eqref{eqn:crit_loc_summary} is consistent with investigations of MBL in classical spin systems~\cite{Oganesyan2009,Basko2011}. As a thermal classical spin presents a continuous spectrum to the adjacent non-chaotic spins, it completely thermalizes a putatively-MBL chain, so that MBL is not stable in classical systems. In our case also, if \(\xi_f\) grows faster than a power law, \(\nu \to \infty\) (which includes the case of the spectrum being continuous at finite \(N\)), the critical localization length is zero.

	In \autoref{subsec:model} we state our model of a thermal inclusion in a quasiperiodically-driven putatively MBL chain. Then in \autoref{subsec:avalanche} and Appendix~\ref{app:chi_star} we derive Eq.~\eqref{eqn:crit_loc_summary}.

	\subsection{Model}
		\label{subsec:model}

		The Hilbert space (in the temporal domain) for the putatively MBL chain is \(\mathcal{H} = \mathcal{H}_B \otimes \mathcal{H}_{\mathrm{MBL}}\), where \(\mathcal{H}_B\) is the \(N\)-dimensional Hilbert space of the thermal inclusion (the ``bath''), and \(\mathcal{H}_{\mathrm{MBL}}\) is the Hilbert space of the MBL chain, which we regard as a tensor product of two-level systems -- the l-bits.

		The Hamiltonian on this system is \(H(t) = H_0(\Bt_t) + H_{\mathrm{int}}(\Bt_t)\), where \(H_0(\Bt)\) consists of the uncoupled Hamiltonians of the thermal region and the MBL chain, and \(H_{\mathrm{int}}(\Bt)\) is the interaction between them.

		In the frequency lattice, we have a quasienergy operator \(\tilde{K} = \tilde{K}_0 + \tilde{K}_{\mathrm{int}}\), with
		\begin{equation}
			\tilde{K}_0 = -\sum_{\Bn} \BO\cdot\Bn \ketbra{\Bn}{\Bn} + \tilde{K}_B \otimes \mathbbm{1}_{\mathrm{MBL}} + \mathbbm{1}_B\otimes \tilde{K}_{\mathrm{MBL}},
		\end{equation}
		where \(\tilde{K}_a\) for \(a \in \{B, \mathrm{MBL}, \mathrm{int}\}\) is a translationally invariant term, and
		\begin{equation}
			\tilde{K}_{\mathrm{int}} = \sum_{j} J_j(\tilde{V}\tilde{\tau}_j^+ + \tilde{V}^\dagger \tilde{\tau}_j^-).
			\label{eqn:Kint}
		\end{equation}
		Here, \(\tilde{V}\) is a not-necessarily-Hermitian operator acting on the bath with \(O(1)\) operator norm, and \(\tilde{\tau}^\pm_j\) are the raising and lowering operators for the l-bit \(\tilde{\tau}^z_j\). Localization of the l-bits implies that the coefficients \(J_j = O(e^{-|j|/\xi_s})\) decay exponentially in space. We have suppressed a dependence on \(j\) from the terms \(\tilde{V}\).

		The assumed form of the interaction~\eqref{eqn:Kint} is incomplete. We have neglected products of l-bit operators, and have not included a term like \(J_j' \tilde{V}' \tilde{\tau}^z_j\) which does not flip l-bits. These additional terms do not change the results of our analysis~\cite{Crowley2021}.

	\subsection{Thermal Avalanches}
		\label{subsec:avalanche}

		\subsubsection{The Fidelity Susceptibility in the Frequency Lattice}
			\label{ssubsec:fid_sus}

			We consider l-bits two at a time -- one on each side of the thermal region, which we position at \(j = 0\) (\autoref{fig:therm_incl}).

			To quantify when the l-bits at \(\pm j\) are thermalized by the thermal inclusion, we will use the \emph{fidelity susceptibility} in the frequency lattice. The fidelity susceptibility \(\chi_j\) quantifies the strength of hybridization between uncoupled eigenstates that differ in the \(\pm j\)th l-bit when said l-bits are coupled to the bath.

			Uncoupled frequency lattice quasienergy states of the thermal region and the two l-bits take the form of a convolution (\autoref{sec:back}),
			\begin{equation}
				\ket{\tilde{\phi}_{\alpha}\tilde{\tau}_{j} \tilde{\tau}_{-j}} = \ket{\tilde{\phi}_{\alpha}}*\ket{\tilde{\tau}_{j}}*\ket{\tilde{\tau}_{-j}},
				\label{eqn:uncoupled_qsts}
			\end{equation}
			with quasienergy
			\begin{equation}
				\epsilon_\alpha + h_j \tau_{j} + h_{-j} \tau_{-j}
			\end{equation}
			where \(\tau_{j},\tau_{-j} = \pm1\) and \(\alpha\) indexes the Hilbert space of the thermal region. (We have neglected products of l-bit operators in \(\tilde{K}_{\mathrm{MBL}}\) by assuming this form of the quasienergy.)

			The fidelity susceptibility can be regarded as the norm of the correction to this state in the first order of perturbation theory, regarding the coupling \(\tilde{K}_{\mathrm{int}}\) as a perturbation,
			\begin{equation}
				\chi_\alpha = \sum_{\beta, \Bl,h} \left|\frac{\tilde{V}_{\beta\alpha}^{\Bl}}{\omega_{\alpha\beta} + \Bl\cdot\BO - 2h}\right|^2.
				\label{eqn:fid_sup}
			\end{equation}
			Here, the sum excludes \((\beta,\Bl) = (\alpha, 0)\), but it includes \(\beta=\alpha\) when \(\Bl \neq 0\). The matrix elements \(\tilde{V}_{\beta\alpha}^{\Bl} = \bra{\tilde{\phi}_\beta^\Bl} \tilde{V} \ket{\tilde{\phi}_\alpha}\) will be taken to be of the form proposed in \eqref{eqn:RMT_ans}. The denominator
			\begin{equation}
				\Delta_{\beta\alpha}^{\Bl} = \omega_{\alpha\beta} + \Bl\cdot\BO - 2h
			\end{equation}
			is the quasienergy difference between the states \(\ket{\tilde{\phi}_\alpha \{\tilde{\tau}\}}\) and \(\ket{\tilde{\phi}_\beta^\Bl \{\tilde{\tau}'\})}\), so that
			\begin{equation}
				h \in \{2h_j \tau_j, 2h_{-j} \tau_{-j} \}
			\end{equation}
			depending on whether l-bit \(j\) or l-bit \(-j\) is flipped by \(\tilde{K}_{\mathrm{int}}\) (at the first order of perturbation theory, only one can be flipped).

			In the static case, the distribution of \(\chi_\alpha\) within a particular random matrix ensemble for \(V\) can be calculated in many cases~\cite{Crowley2021}. In all cases, it has a broad distribution with a power-law tail. As we show in Appendix~\ref{app:chi_star}, this is also true in the frequency lattice.

			That is, \(\chi_\alpha\) has a distribution function with asymptotic behavior,
			\begin{equation}
				f_{\mathrm{FS}}(\chi) \overset{\chi\to\infty}{\sim} \sqrt{\frac{\chi_{\star,j}}{\chi^3}},
			\end{equation}
			where the typical scale of the distribution is given by 
			\begin{equation}
				\sqrt{\chi_{\star,j}} = \lim_{\Delta \to 0}\left[\frac{1}{2\Delta}\sum_{|\Delta_{\beta\alpha}^{\Bl}|<\Delta} |\tilde{V}^{\Bl}_{\beta\alpha}|\right].
				\label{eqn:chi_star}
			\end{equation}
			Here, the sum is over all uncoupled quasienergy states \(\ket{\tilde{\phi}_\beta^\Bl \{\tilde{\tau}'\})}\) such that \(|\Delta_{\beta\alpha}^{\Bl}|<\Delta\), and the square brackets indicate an average over the distribution of matrix elements \(|\tilde{V}^{\Bl}_{\beta\alpha}|\) (determined by the distribution of the random numbers \(R_{\beta\alpha,\Bl}\) of \eqref{eqn:RMT_ans}) and of the level spacings \(|\Delta_{\beta\alpha}^{\Bl}|\). We have not been specific about the distributions for the matrix element or level spacing for two reasons: first, \(\sqrt{\chi_{\star,j}}\) is well-defined given only very weak conditions on the distributions (the probability density that \(\Delta_{\beta\alpha}^{\Bl}=0\) is finite, and the averages of \(|\tilde{V}^{\Bl}_{\beta\alpha}|\) are summable over \(\Bl\)); and secondly, we will only be concerned with the scaling properties of \(\sqrt{\chi_{\star,j}}\). To calculate the actual value of \(\sqrt{\chi_{\star,j}}\) we would need these distributions, but they are unnecessary to deduce the fidelity susceptibility's asymptotic behavior with \(j\).

			The dimensionless quantity \(J_j\sqrt{\chi_{\star,j}}\) formalizes the notion of the product of a matrix element and a density of states in the frequency lattice.

		\subsubsection{Growth of the Thermal Region}
			\label{ssubsec:growth_therm}

			If the spin chain is thermal, the dimensionless combination \(J_j \sqrt{\chi_j}\) remains large as \(j \to \infty\) -- all uncoupled quasienergy states hybridize strongly to form highly entangled thermal eigenstates. In an MBL system, \(J_j \sqrt{\chi_j}\) decreases to zero, indicating that l-bits distant from the thermal inclusion are only slightly dressed by their coupling to said inclusion. Our aim is to show that the latter scenario of \(J_j \sqrt{\chi_j} \to 0\) is possible. In this stage, we mimic the arguments of Ref.~\cite{deRoeck2017}.

			We make the pessimistic assumption that all l-bits up to and including \(\pm|j-1|\) are perfectly absorbed by the thermal region. Then the system consisting of the thermal region and the first \(2(j-1)\) l-bits is still described by the random matrix ansatz \eqref{eqn:RMT_ans}, but with a larger Hilbert space dimension \(N_{j-1} = N 2^{2(j-1)}\). This in turn generically affects the frequency lattice localization length \(\xi_f = \xi_f(N_{j-1})\), and subsequently affects the spectral functions \(|f_\Bl| = O(e^{-|\Bl|/\xi_f(N_{j-1})})\).

			Considering some fixed quasienergy window \(\Delta\), there are on the order of
			\begin{equation}
				N_{\mathrm{eff},j-1} = N_{j-1} \xi_f(N_{j-1})^{D-1}
			\end{equation}
			terms in the sum that contribute to \(\sqrt{\chi_{\star,j}}\) before the exponential suppression from \(|f_\Bl|\) makes further terms negligible. (Recall from \autoref{sec:ansatz} that the relevant frequency lattice volume is \(\xi_f^{D-1}\), and not \(\xi_f^D\), because the extent of the quasienergy state in the direction parallel to \(\BO\) scales weakly, i.e. slower than a power law, with \(N\).)

			Meanwhile, each of the terms within the localization length scales as \(N_{\mathrm{eff},j-1}^{-1/2}\), in order to fix the normalization of \(V\). Thus, we see that
			\begin{equation}
				\sqrt{\chi_{\star,j}} = O\left(\sqrt{N_{\mathrm{eff},j-1}}\right).
			\end{equation}

			By assumption of MBL, we have that \(J_j = O(e^{-|j|/\xi_s})\), so for the dimensionless quantity \(J_j \sqrt{\chi_{\star,j}}\), we have,
			\begin{equation}
				\log(J_j \sqrt{\chi_{\star,j}}) = O\left( -\frac{j}{\xi_s}+ j\log 2 + \frac{D-1}{2}\log\frac{\xi_f(N 2^{2j})}{\xi_f(N)} \right).
			\end{equation}
			The thermal avalanche will eventually stop if 
			\begin{equation}
				\lim_{j \to \infty} \left[-\frac{j}{\xi_s}+ j\log 2 + \frac{D-1}{2}\log\frac{\xi_f(N 2^{2j})}{\xi_f(N)}\right] = -\infty.
			\end{equation}
			This requires that the frequency lattice localization length grows at most as a power law in the Hilbert space dimension of the bath,
			\begin{equation}
				\xi_f(N) = O(N^\nu),
				\label{eqn:xi_scale}
			\end{equation}
			that is, at most exponentially in the number of thermal spins.

			Assuming~\eqref{eqn:xi_scale}, there is a critical spatial localization length \(\xi_{s,c}\) below which MBL is stable, just as in the case of static MBL. In the quasiperiodically driven case, this is given by
			\begin{equation}
				\xi_{s,c}^{-1} = [1 + \nu(D-1)]\log 2.
				\label{eqn:crit_loc}
			\end{equation}
			For spatial localization lengths below this value, \(\xi_s < \xi_{s,c}\), the susceptibility \(J_j \sqrt{\chi_{\star,j}}\) decreases exponentially with \(j\). Otherwise, the thermal region grows to encompass the entire system.

			We note that the result \eqref{eqn:crit_loc} has Floquet MBL as a special case with \(D=1\). In that case, the critical localization length is the same as the static case, \((\log 2)^{-1}\), as is already well known from other arguments based on the Floquet Hamiltonian~\cite{Ponte2015,Abanin2016,Crowley2020b}.

			We also observe that the quasiperiodically driven MBL phase is less stable than the static phase, in the sense that the critical localization length is strictly smaller than that in the static case. This is because the presence of the frequency lattice allows the effective Hilbert space dimension \(N_{\mathrm{eff}}\) to grow faster than \(2^{2j}\).

\section{Synthetic Localization for Two-Tone Driving}
	\label{sec:synth}

	In this section we show that \eqref{eqn:xi_scale} generically holds for \(D=2\) with \(\nu=1\), and thus that quasiperiodically-driven MBL is stable to thermal inclusions in the case of two-tone driving, with a critical localization length
	\begin{equation}
		\xi_{s,c} = (2\log 2)^{-1}.
	\end{equation}

	The localization of quasienergy states for smooth two-tone driving can be understood as Anderson localization in the \(D-1=1\) dimensional surface perpendicular to \(\BO\) in the frequency lattice. That is, it is essentially a \emph{single-particle} effect, even in this many-body setting.

	Note that the localization is ``generic'' -- there are finely-tuned examples in the literature of smooth two-tone driving resulting in delocalized quasienergy states~\cite{Blekher1992,Crowley2019}.

	For \(D=2\), Stark localization produces a quasi-one-dimensional model of width roughly \(\xi_{\parallel} \approx W/|\BO|\) along which quasienergy states could delocalize, where \(W\) is the bandwidth of the static part of the Hamiltonian. We lump together sites along the width of this strip to form new sites with increased Hilbert space dimension \(N' \approx N W/|\BO|\) and bandwidth \(W' \approx 2W\). In this coarse-grained model, localization is nearly complete in the direction parallel to \(\BO\). We drop the primes on \(N'\) and \(W'\), and consider the one-dimensional model thus formed below.

\begin{figure}
		\centering
		\includegraphics[width=\linewidth]{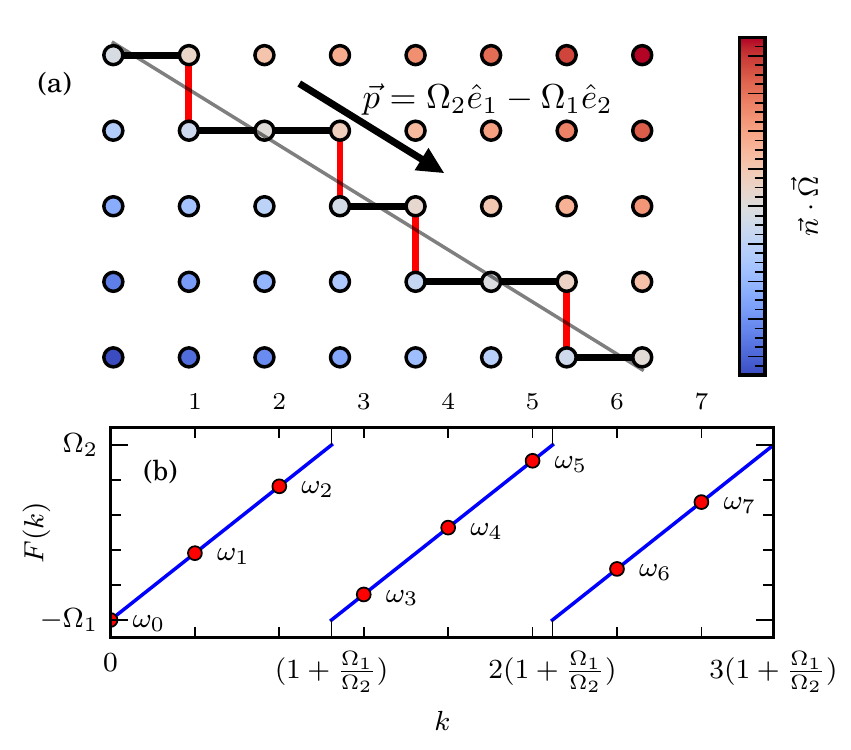}
		\caption{\label{fig:quasi1d}\emph{Quasi-one-dimensional model.}--- (\textbf{a})~Restricting the full two-dimensional (coarse-grained) frequency lattice to those sites closest to a given equipotential (grey line) with tangent \(\vec{p}\) produces a one-dimensional model~\eqref{eqn:K1dim}. The model has a quasiperiodic potential and distinct hopping matrices on the horizontal and vertical bonds. (\textbf{b})~The on-site potentials \(\omega_k = \Bn_k\cdot\BO\) for sites \(\Bn_k\) in the one-dimensional model are quasiperiodic. They are obtained by sampling a sawtooth function \(F(k)\) incommensurately to its period of \(1+\Omega_1/\Omega_2\). The discontinuity in \(F\) favors localization in the one-dimensional model.}
	\end{figure}

	The sequence of sites included in the one-dimensional model are those closest to the line with tangent \(\vec{p} = \Omega_2 \hat{e}_1 - \Omega_1 \hat{e}_2\)~(\autoref{fig:quasi1d}). Label these sites by the index \(k\) such that \(\Bn_k = \Bn_{k-1} \pm \hat{e}_{i_k}\), where the sign of \(\pm \hat{e}_{i_k}\) is determined by the sign of \(\Omega_1\) and \(\Omega_2\), and \(i_k \in \{1,2\}\) is a sequence determined by the number theoretic properties of \(\Omega_1/\Omega_2\). For instance, when \(\Omega_1/\Omega_2 = (1+\sqrt{5})/2\) is the golden ratio, \(i_k\) is the Fibonacci word of the elements \(\{1,2\}\)~\cite[Chapter 2]{Brown1993,Lothaire2002}.

	The quasienergy states are then approximated by the mid-spectrum eigenstates of the one-dimensional single-particle tight-binding model~\footnote{The states in the middle of the spectrum have localization centers in the middle of the coarse-grained strip, and are thus least affected by the truncation to a one-dimensional model.}
	\begin{equation}
		K_{\text{1-dim}} = \sum_{k,k' \in \integers} (H_{kk'} - \omega_k \delta_{kk'}) \ketbra{k}{k'},
		\label{eqn:K1dim}
	\end{equation}
	with \(N\) orbitals per site and where \(H_{kk'} = H_{\Bn_k - \Bn_{k'}}\) still decays exponentially in \(|k-k'|\), but is not necessarily translationally invariant in \(k\). The on-site potential \(\omega_k = \Bn_k \cdot \BO\) is defined up to a constant by the recursion
	\begin{equation}
		\omega_{k} = \omega_{k-1} + (-1)^{i_k}\Omega_{i_k},
		\label{eqn:omega_recursion}
	\end{equation}
	where we have chosen \(k\) to increase in the direction of \(\vec{p} = \Omega_2 \hat{e}_1 - \Omega_1 \hat{e}_2\).

	The potentials \(\omega_k\) are quasiperiodic in the sense that they may be obtained by sampling a periodic function \(F(k)\) at a rate incommensurate to the period of \(F\). Indeed, one can check that taking
	\begin{equation}
		F(k+\beta) = F(k) = \Omega_2 k + C \quad \text{for } k \in [0,\beta)
		\label{eqn:F}
	\end{equation}
	as piecewise linear with period \(\beta = 1 + \Omega_1/\Omega_2\) (so that \(F\) is a sawtooth, \autoref{fig:quasi1d}(b)) recovers \(F(k) - F(k-1) = (-1)^{i_k}\Omega_{i_k}\).

	The Hamiltonian \eqref{eqn:K1dim} is an inhomogeneous one-dimensional hopping problem. Such a model has exponentially localized eigenstates if the on-site potential is random and the hopping is quasilocal~\cite{Anderson1958}. Although the potentials \(\omega_k\) in~\eqref{eqn:omega_recursion} are not random, we argue that the intuition from Anderson localization is correct in this case, and that the localization of the model~\eqref{eqn:K1dim} is captured by the associated Anderson model
	\begin{equation}
		K_{\text{random}} = \sum_{k,k' \in \integers} (H_{kk'} - \omega'_k \delta_{kk'}) \ketbra{k}{k'},
		\label{eqn:K1dim_random}
	\end{equation}
	where \(\omega'_k\) are independent random variables sampled from the uniform distribution on \([C, C+ \Omega_1 + \Omega_2)\).

	\subsection{Localization in the Anderson Model}
		\label{subsec:loc_anderson}

		The localization of the Anderson chain \(K_{\text{random}}\) is controlled by the ratio \(r\) of typical hopping amplitudes to the scale of the disorder. By estimating \(r\), we obtain a prediction for the dependence of the localization length of the quasienergy states \(\zeta_f\) (and hence that for the matrix elements, \(\xi_f = O(\zeta_f)\)), on the number of orbitals \(N\).

		We begin by estimating the effective scale of the disorder in the \(N\)-band model Eq.~\eqref{eqn:K1dim_random}. A quasienergy state with quasienergy \(\epsilon_0\) in the uncoupled model, with \(H_{kk'}\) set to zero for \(k \neq k'\), will hybridize with states with a similar quasienergy. This justifies considering the delocalization of this state as only involving the energy levels on each site closest to \(\epsilon_0\). The uncoupled energy levels of \(H_0 = H_{kk}\) have a typical density of states in the middle of the spectrum given by \(\rho = N/W\). If \(H_0\) is modeling a many-body Hamiltonian on \(L\gg 1\) spins, then \(W = O(\sqrt{L})\), and \(|\BO| \ll W\) at large \(L\) -- the on-site potential is small compared to the bandwidth. Then we can approximate the density of states at quasienergy \(\epsilon_0\) on every other site in the chain as also being \(\rho = N/W\). The effective disorder strength in the Anderson model~\eqref{eqn:K1dim_random} is thus set by the typical level spacing between these states: \(W/N\).

		If the hopping matrices \(H_{kk'}\) have typical scale \(\|H_{kk'}\| = J\), where \(J\) depends on the driving protocol, then the typical scale of the matrix element between the resonant levels is \(J/\sqrt{N}\). The factor of \(\sqrt{N}\) comes from an assumption that the eigenstates of \(H_0\) present themselves in matrix elements of \(H_{kk'}\) as random vectors~\cite{DAlessio2016}.

		The hopping \(J/\sqrt{N}\) is asymptotically larger than the ``disorder'' \(W/N\), so as \(N \to \infty\) the model~\eqref{eqn:K1dim_random} must enter the low-disorder regime. Indeed,
		\begin{equation}
			r = \frac{J/\sqrt{N}}{W/N} = (J/W)\sqrt{N}.
			\label{eqn:hop_dis_ratio}
		\end{equation}
		grows without bound with \(N\).

		In the large \(r\) regime, the localization length scales with \(r^2\)~\cite{Thouless1973}, giving
		\begin{equation}
			\xi_f = O(r^2) = O\left(\left(\tfrac{J}{W}\right)^2 N\right).
		\end{equation}
		This provides \(\nu = 1\). Not only is \(\xi_f\) finite for all finite \(N\), it grows only linearly with \(N\) (that is, as \(2^L\) in a spin chain).

	\subsection{Localization in the Quasiperiodic Model}
		\label{subsec:loc_quasiperiodic}

		While the inhomogeneous on-site potentials \(\omega_k\) in the model~\eqref{eqn:K1dim} are not random, we find the associated Anderson model~\eqref{eqn:K1dim_random} to be an effective description of the localization properties of the system. This can be verified numerically, and partially justified analytically.

\begin{figure}
			\centering
			\includegraphics[width=\linewidth]{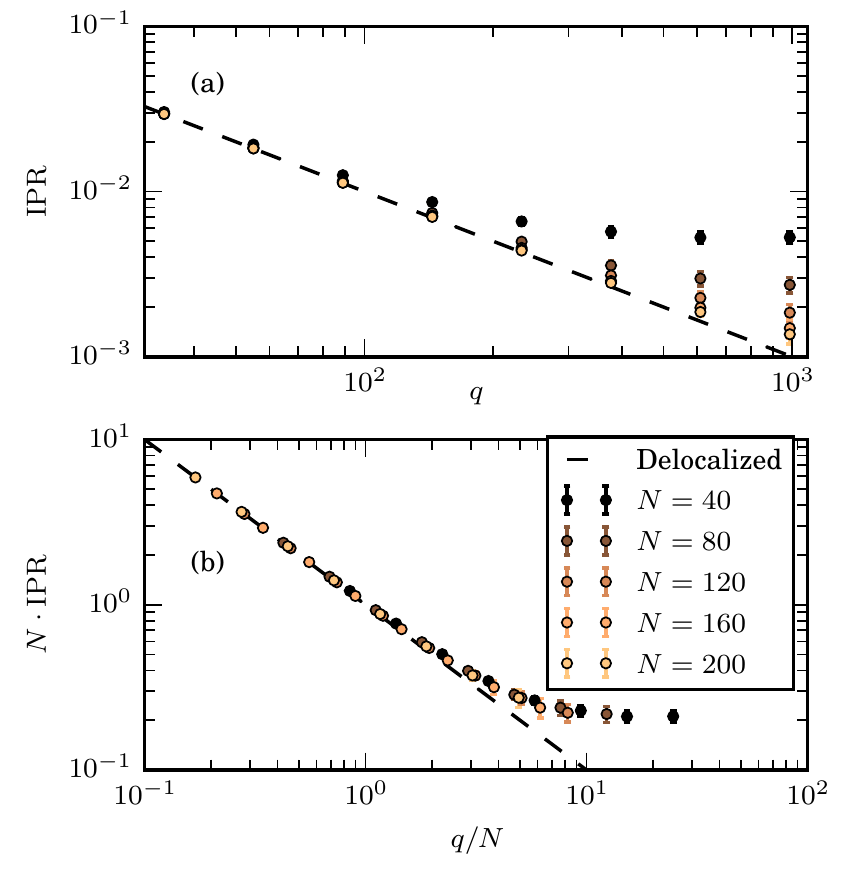}
			\caption{\label{fig:commen_IPR}\emph{IPR in commensurate approximations.}--- (\textbf{a})~Inverse participation ratio averaged over both eigenstate index \(\alpha\) and \(N_{\mathrm{samp}} \approx 1200/N\) samples in the random matrix model of Appendix~\ref{subsec:model_synth}. Successive commensurate approximations are indexed by Fibonacci numbers \(q = F_n\). For small \(q\), the \(\mathrm{IPR}\) decreases as \(1/q\) (dashed line), but for large \(q\) the \(\mathrm{IPR}\) saturates, indicating a finite localization length in the incommensurate limit. (\textbf{b})~Scaling by \(N\) leads to a good data collapse, consistent with \(\nu = 1\). \emph{Parameters for model~\eqref{eqn:RMT_model2}:}  \(J/W = 0.2\), \(\Omega_1/W = 0.6\), \(q \in \{34,\ldots,987\}\), with \(N_{\mathrm{samp}} \approx 1200/N\) random matrix samples.}
		\end{figure}

		The prediction of exponential localization with \(\nu=1\) can be checked numerically in a driven random matrix model. Detailed descriptions of these numerics can be found in Appendix~\ref{app:numerics_synth_loc}, but we summarize some findings here. By taking a commensurate approximation to \(\BO\) it is possible to calculate quasienergy states (Appendix~\ref{subsec:comens}). \autoref{fig:commen_IPR} shows the \emph{inverse participation ratio} (IPR)~\eqref{eqn:IPR} of the quasienergy states in a series of commensurate approximations indexed by Fibonacci numbers \(q = F_n\). The IPR is roughly \(\mathrm{IPR} \propto \xi_f^{-1}\), so seeing the IPR saturate as \(q\to\infty\) indicates the localization length is finite in the incommensurate limit. Rescaling the IPR by \(N\) and \(q\) by \(1/N\) produces a good data collapse, consistent with \(\nu = 1\).

		The one-dimensional model~\eqref{eqn:K1dim} can also be simulated directly (Appendix~\ref{subsec:quasi1d}). This produces a more quantitative prediction that
		\begin{equation}
			\nu = 1.001 \pm 0.009,
		\end{equation}
		which is also consistent with \(\nu=1\).

		The model~\eqref{eqn:K1dim} evades the mechanism of delocalization in many well-known quasiperiodic models, such as the Aubry-Andr\'e model~\cite{Aubry1980,Harper1955}. Determining the localization properties of quasiperiodic tight-binding models, such as Eq.~\eqref{eqn:K1dim}, is assisted by the existence of a duality transformation of these models~\cite{Aubry1980,Harper1955,Hiramoto1992,Han1994,Sarang2017,Chandran2017,Crowley2018}. For simplicity, suppose the hopping matrices \(H_{kk'} = H_{k-k'}\) are translationally invariant (which amounts to an isotropy condition in the two-dimensional frequency lattice: \(H_{\hat{e}_1} = H_{\hat{e}_2}\), etc.). Then the dual model is related to Eq.~\eqref{eqn:K1dim} by Fourier transform. Indeed, if \(\ket{\tilde{\phi}} = \sum_k \ket{\phi_k}\ket{k}\) is an eigenstate of \(K_{\text{1-dim}}\), then substituting the Fourier transform
		\begin{equation}
			\ket{\phi_k} = C \sum_{x} \ket{\phi_x} e^{-2\pi i x k/\beta}
		\end{equation}
		(where \(C\) is a normalization constant) into the eigenvalue equation \(K_{\text{1-dim}}\ket{\tilde{\phi}} = \epsilon \ket{\tilde{\phi}}\) reveals that \(\sum_x \ket{\phi_x}\ket{x}\) is an eigenstate of 
		\begin{equation}
			K_{\text{1-dim}}^{\mathrm{dual}} = \sum_{x,x' \in \integers} (H(x) \delta_{x x'} - F_{x - x'} ) \ketbra{x}{x'},
			\label{eqn:K1dim_dual}
		\end{equation}
		where
		\begin{align}
			H_k &= C\sum_{x} H(x) e^{2\pi i x k/\beta}, \\
			F(k) &= C\sum_{x} F_x e^{-2\pi i x k/\beta}.
		\end{align}
		
		If an eigenstate \(\ket{\tilde{\phi}}\) of \(K_{\text{1-dim}}^{\mathrm{dual}}\) is localized, then the dual eigenstate of \(K_{\text{1-dim}}\) must be delocalized~\cite{Crowley2018}. In the self-dual Aubry-Andr\'e model~\cite{Aubry1980,Harper1955}, this guarantees the existence of a delocalized phase. Similarly, whenever the on-site potential \(F(k)\) is smooth and the hopping amplitudes \(H_{k-k'}\) decay exponentially, the dual model also has a smooth potential and exponentially decaying hopping amplitudes. At least one of the two models related by duality must be delocalized, and as both models have a similar structure, it is not possible for quasiperiodic models with smooth potentials to generically be localized.

		In contrast, we observe that the on-site potential \(F(k)\) in Eq.~\eqref{eqn:F} is not smooth as a function of \(k\) -- it has a finite jump -- and so the hops \(F_x\) in the dual model are power-law decaying. In the absence of other special structure, we expect that the long-range model \(K_{\text{1-dim}}^{\mathrm{dual}}\) will be delocalized, which allows the quasilocal hopping model \(K_{\text{1-dim}}\) to generically be localized.
		
		While it is possible for both \(K_{\text{1-dim}}\) and \(K_{\text{1-dim}}^{\mathrm{dual}}\) to be delocalized, once the inhomogenous model \(K_{\text{1-dim}}\) evades any condition preventing it from localizing, the intuition from Anderson localization is that it will do so. Our numerical results provide a strong case for generic localization with \(\nu=1\).

\section{Absence of Synthetic Localization with Three or More Tones}
	\label{sec:more_tones}

	Following \autoref{sec:synth}, Stark localization produces a coarse-grained single-particle hopping problem in \(D-1\) dimensions. Just as in the \(D=2\) case, said hopping model has a large number of orbitals \(N\), an inhomogeneous on-site potential \(\omega_{\Bk} = \BO\cdot\Bn_\Bk\) and exponentially decaying (but no longer necessarily translationally invariant) hopping matrices \(H_{\Bk\Bk'}\).

	A disordered \((D-1)\)-dimensional Anderson model with spin-orbit coupling is not always localized  for \(D\geq 3\)~\cite{Abrahams1979,Hikami1980,Altshuler1980}. There is typically a non-zero hopping amplitude to disorder strength ratio \(r_c\) above which eigenstates become delocalized. As argued in \autoref{sec:synth}, the relevant ratio in our case is \(r = (J/W)\sqrt{N}\)~\eqref{eqn:hop_dis_ratio}.

	The ratio \(r\) grows with \(N\), so for sufficiently large \(N>N_c\), the ratio \(r\) exceeds the critical value \(r_c\), and the localization length \(\xi_f\) becomes infinite. That is, a large enough, but finite, thermal inclusion acquires a genuinely continuous spectrum in the presence of three-tone (or more) driving. This feature destabilizes randomly disordered MBL -- it is known from, for instance, Ref.~\cite{Oganesyan2009} that a finite thermal region that presents a continuous spectrum to the rest of the chain can completely thermalize the system given sufficient time.

	That \(\xi_f\) can diverge for a finite thermal inclusion is supported by recent numerical evidence. Refs.~\cite{Long2021,Nathan2020b} identify a phase believed to be delocalized in the frequency lattice for a three-tone-driven qubit (\(N=2\)). Larger \(N\) only increases the likelihood to delocalize.

	We conclude that quasiperiodically-driven MBL with random disorder can only be stable for two-tone-driving. For \(D \geq 3\) tones, sufficiently large thermal regions will destabilize a putatively MBL chain.

\section{Many-Body Resonances}
	\label{sec:many_res}

	Another mechanism for destroying MBL is the proliferation of \emph{many-body resonances} -- if, for all \(L\) sufficiently large, a fixed nonzero perturbation to a putatively MBL chain causes a given quasienergy state to hybridize with exponentially many in \(L\) other quasienergy states, then MBL is not a stable dynamical phase.

	In static systems, demanding perturbative stability of MBL implies that the localization length must be below a critical value \(\xi'_{s,c}\). This critical value is bounded from below by \((\log 2)^{-1}\), the critical localization length predicted by the avalanche argument. The bound is saturated when the matrix elements of the perturbation between l-bit states that differ in \(\tilde{\tau}^z_{\pm r}\) are sufficiently narrowly distributed. In physical chains, however, the matrix elements at each range \(r\) are broadly distributed (approximately log-normally)~\cite{Varma2019}, so that there is a window of disorder strengths accessible at small sizes in which localization in the chain is stable to the formation of many-body resonances, but not to thermal avalanches~\cite{Crowley2020b,Morningstar2021}.

	For \(D \geq 2\) tones, we show that the critical localization length for perturbative stability is still bounded by \((\log 2)^{-1}\), which is now strictly larger than the localization length provided by the avalanche argument. Thus, we expect that the regime wherein avalanches, and not many-body resonances, control the (in)stability of randomly disordered MBL (\(\xi_{s,c} < \xi_s < \xi'_{s,c}\)) is broader in quasiperiodically-driven systems than in static and periodically-driven systems.

	The reason the bound on \(\xi'_{s,c}\) is unaltered from the static case is because the frequency lattice only provides a polynomial enhancement to the effective density of states introduced in \autoref{sec:space}. Unlike in the case of a thermal avalanche, there is no growing thermal bubble that can expand exponentially in the frequency lattice as it absorbs more spins. Without the required exponential scaling, the effective density of states cannot compete with the decaying matrix elements. The remainder of this section is essentially a formal verification of this intuition.

	The frequency lattice fidelity susceptibility (\autoref{ssubsec:fid_sus}) detects if a perturbation \(J V(\Bt)\) to a putatively MBL Hamiltonian \(H(\Bt)\) causes large changes to the unperturbed quasienergy states. Strong localization of l-bits places constraints on the fidelity susceptibility, and ensures perturbative stability. This calculation generalizes methods used in Ref.~\cite{Crowley2020b} in the static and Floquet contexts.

	We assume that the Hermitian operator \(V(\Bt)\) is quasilocal in space centered at \(j=0\) (say), and smooth in \(\Bt\). To extract the spatial structure of \(\tilde{V}\) it is convenient to decompose it as
	\begin{equation}
		\tilde{V} = \sum_r \tilde{V}_r
	\end{equation}
	where \([\tilde{V}_r, \tilde{\tau}^z_j] = 0\) for \(|j| > r\), and \([\tilde{V}_r, \tilde{\tau}^z_{\pm r}] \neq 0\). In words, \(\tilde{V}_r\) acts trivially on l-bits that are further than a range \(r\) from \(j=0\), and non-trivially on those exactly at range \(r\).
	We define a scaled Frobenius norm for the temporal operator for \(\tilde{V}_r\),
	\begin{equation}
		\|V_r\| = \sqrt{\frac{1}{2^L}\int \frac{\d^D\theta}{(2\pi)^D} \tr{V_r(\Bt)^2}}
		\label{eqn:norm}
	\end{equation}
	where \(L\) is the system size. Quasilocality of \(\tilde{V}\) in real space is expressed as
	\begin{equation}
		\log \|V_r\| \sim -\frac{r}{\xi_s}.
	\end{equation}
	Quasilocality in the synthetic dimensions implies exponential decay of the matrix elements \(\tilde{V}^{\Bl}_{\beta\alpha}\) with \(|\Bl|\), with localization length \(\xi_f\), as usual.
	
	We use the assumed exponential decay of \(\|V_r\|\) with \(r\) to deduce the scaling of the matrix elements appearing in the calculation of the fidelity susceptibility \(\chi_{\star}\). In terms of the matrix elements between quasienergy states \(\ket{\tilde{\phi}_{\alpha}^\Bn} = \ket{\{\tilde{\tau}\}^\Bn}\), specified by their l-bit configurations and a translation \(\Bn\), the norm is
	\begin{equation}
		\|V_r\|^2 = \frac{1}{2^L}\sum_{\alpha,\beta,\Bl} |(\tilde{V}_r)^{\Bl}_{\beta \alpha}|^2.
		\label{eqn:norm_frq}
	\end{equation}
	
	To estimate \(\chi_\star\), we find the average squared matrix element, summed over \(\Bl\):
	\begin{equation}
		v(r)^2 = \left[ \sum_{\Bl} |(\tilde{V}_r)^{\Bl}_{\beta \alpha}|^2 \right],
	\end{equation}
	where square brackets indicate an average over those \(\alpha\) and \(\beta\) such that the matrix element is non-zero. Comparing this to Eqs.~\eqref{eqn:norm} and~\eqref{eqn:norm_frq}, and noting that there are \(N_r = O(2^{2r})\) states \(\ket{\tilde{\phi}_{\beta}}\) for which the matrix element is non-zero with a given \(\alpha\), we have
	\begin{equation}
		v(r) = \frac{O(e^{-r/\xi_s})}{\sqrt{N_r}}.
		\label{eqn:avg_mat}
	\end{equation}

	By summing over the frequency lattice before analyzing the scaling of \(\chi_\star\), the problem of calculating \(\xi'_{s,c}\) is essentially reduced to the static case. The sum was possible due to the exponential decay of \(|(\tilde{V}_r)^{\Bl}_{\beta \alpha}|^2\) with \(|\Bl|\), and the fact that there are only polynomially many frequency lattice sites with a given \(|\Bl|\).

	Summarizing the remaining steps in the calculation~\cite{Crowley2020b}: one organizes the sum for \(\chi_\star\) in terms of operators of increasing range \(\tilde{V}_r\), which gives
	\begin{equation}
		\sqrt{\chi_\star} \leq \sum_{r=1}^L \left(\lim_{\Delta \to 0}\left[\frac{1}{2\Delta}\sum_{|\Delta^{\Bl}_{\beta \alpha}|<\Delta} |(\tilde{V}_r)^{\Bl}_{\beta \alpha}|\right]\right),
	\end{equation}
	where we used the triangle inequality.
	Eq.~\eqref{eqn:norm_frq} places a restriction on the sum of squares of the matrix elements. Given this restriction, the sum of the absolute values appearing in \(\sqrt{\chi_\star}\) is maximal when all the matrix elements are equal. Thus, we obtain an upper bound for \(\sqrt{\chi_\star}\) by replacing the sum of matrix elements for each \(r\) by the root-mean-square value \(v(r) = O(e^{-r/\xi_s}2^{-r})\) times the number of terms \(N_r = O(2^{2r})\).

	The \(\Delta \to 0\) limit introduces an unimportant \(O(1)\) factor. Thus, we have
	\begin{equation}
		\sqrt{\chi_\star} \leq O\left(\sum_{r=1}^L 2^{r} e^{-r/\xi_s} \right).
	\end{equation}

	Demanding that \(\sqrt{\chi_\star}\) converges as \(L \to \infty\) for \(\xi_s < \xi'_{s,c}\) implies
	\begin{equation}
		\xi'_{s,c} \geq (\log 2)^{-1}.
		\label{eqn:xi_prime}
	\end{equation}
	If this condition is met, then by choosing \(J \ll 1/\sqrt{\chi_\star}\) we have that the dimensionless quantity \(J\sqrt{\chi_\star} \ll 1\), and distant quasienergy states typically do not strongly hybridize when the perturbation \(J V(\Bt)\) is added to the Hamiltonian. On the other hand, if the sum for \(\sqrt{\chi_\star}\) diverges, then no such \(J\) exists in the thermodynamic limit, and the MBL phenomenology is unstable to an arbitrarily small perturbation.

	We conclude that MBL phenomenology is stable to many-body resonances for any number of tones \(D\) whenever the spatial localization length is below a critical value \(\xi'_{s,c} \geq (\log 2)^{-1}\), the bound for which is independent of \(D\).

	We reiterate that Eq.~\eqref{eqn:xi_prime} is \emph{not} the critical localization length for the stability of randomly disordered MBL in the thermodynamic limit. Avalanches are the dominant instability for MBL, and this is particularly stark in quasiperiodically driven MBL for \(D\geq 3\).

\section{Discussion}
	\label{sec:disc}

	We have shown that two-tone-driven randomly-disordered MBL is stable to the occurrence of a large thermal region, and to the addition of a small perturbation to the Hamiltonian. Stability requires that the spatial localization length is below a critical value \(\xi_{s,c} = (2 \log 2)^{-1}\). With three or more tones, however, putative MBL is always unstable to thermal avalanches.

	An immediate consequence of our result is that the two-tone-driven topological orders identified in Refs.~\cite{Else2019,Friedman2020,Long2021,Nathan2020b} have infinite lifetime with sufficient disorder. That is, they characterize genuine dynamical phases of matter.

	We have not \emph{proven} the existence of quasiperiodically-driven MBL. Rather, we have checked for the stability of putative MBL to two particular mechanisms of thermalization that are believed to be the dominant ones in the thermodynamic limit. There has been a recent debate about the existence of MBL even in static systems~\cite{Suntajs2020,Sels2020,Sels2021}. Quasiperiodically-driven MBL is not immune to that debate -- all objections to static MBL apply just as much to quasiperiodically-driven MBL.

	Our results also clarify how quasiperiodic driving enhances the effective Hilbert space dimension of a finite system. This feature could be used to increase the thermalizing ability of small quantum systems, and thus aid in experimental tests of thermalization in nearly-isolated quantum systems~\cite{Kucsko2018,Leonard2020}.

	While we have kept our discussion to smooth driving, our results may hold for continuous and piecewise smooth, but non-analytic drives. Non-analyticities result in power-law hops in the frequency lattice, \(\tilde{V}_{\alpha\beta}^{\Bl} = O(|\Bl|^{-p})\). Our conditions on the drive ensure \(p>1\), so that the analogous Anderson model in the frequency lattice is localized for \(D=2\)~\cite{Anderson1958}, and \(\chi_{\star,j=0}\) is finite. Stability to avalanches additionally requires that \(\chi_{\star,j}\) grows at most exponentially in \(j\). We expect this is so, but we leave this calculation to future work.

	Resonance counting in the frequency lattice suggests that discontinuous two-tone drives with \(p<1\) lead to delocalization~\cite{Anderson1958}. Indeed, this has been shown for specific drives in a two-level system~\cite{Luck1988}. As local regions have continuous spectra, MBL is not stable here, explaining the results in Ref.~\cite{Dumitrescu2018}. The marginal \(p=1\) case is an interesting topic for future research~\cite{Levitov1999}.

	With \(D\geq 2\), we expect that the finite size regime in which a localized chain is stable to many-body resonances but not to thermal avalanches is broader than in static and periodically driven systems~\cite{Morningstar2021}. Quasiperiodic driving may thus provide a good experimental setting for the controlled exploration of different instabilities of randomly disordered MBL~\cite{Schreiber2015,Smith2016,Bordia2017,Leonard2020}.

	If putative MBL is due to quasiperiodic spatial modulation (QPMBL), rather than random disorder, then regions of low disorder do not occur~\cite{Iyer2013,Luschen2017}. Avalanches can only occur for \(\xi_s > \xi_{s,c}\) if there is some other mechanism to generate large thermal subsystems. This leaves open the possibility that quasiperiodically-driven QPMBL, and the associated topological dynamical phases~\cite{Else2019,Long2021}, are stable with any number of tones \(D\).

	Our discussion of the critical localization length \(\xi_{s,c}\) largely follows Reference~\cite{deRoeck2017}. Ref.~\cite{deRoeck2017} identifies a bare localization length that is subject to a renormalization group (RG) scaling~\cite{Zhang2016,Thiery2018,Goremykina2019,Morningstar2020}. The value \(\xi_{s,c} = (2 \log 2)^{-1}\) should also be interpreted in this way. We leave a more systematic formulation of RG in quasiperiodically-driven MBL to future work.

	Local integrals of motion could be explicitly constructed on the frequency lattice (\autoref{sec:QPDMBL}) by adapting existing analytical and numerical techniques for static systems~\cite{Ros2015,Chandran2015,Rademaker2016,Imbrie2016,Pekker2017,Kulshreshtha2018}. We suspect the frequency lattice also provides a formalism to generalize Imbrie's proof of static MBL~\cite{Imbrie2016}.

	Our quasiperiodically-driven ETH-style ansatz~\eqref{eqn:RMT_ans} is appropriate for systems with pure-point spectra: finite systems with quasienergy states localized in the synthetic dimensions of the frequency lattice. With three or more tones, the quasienergy states may be delocalized. The ansatz~\eqref{eqn:RMT_ans} can be adapted to this case by taking commensurate approximations to \(\BO\). This collapses the frequency lattice into a cylinder with a finite circumference~\cite{Martin2017,Crowley2019,Long2021}. Quasienergy states are localized parallel to the length of the cylinder by the electric field \(\BO\), but are delocalized around the circumference. The appropriate ETH-style ansatz then becomes
	\begin{equation}
		\tilde{V}_{\alpha \beta}^{\Bl} = \bar{V}_{\Bl} \delta_{\alpha \beta} + \frac{f(\Delta^{\Bl}_{\alpha \beta})}{\sqrt{N \mu}} R_{\alpha \beta,\Bl},
		\label{eqn:RMT_commens_ans}
	\end{equation}
	where symbols are defined as in Eq.~\eqref{eqn:RMT_ans}, and \(\mu\) is the \((D-1)\)-dimensional volume of the cylinder section perpendicular to \(\BO\). Note that the spectral function \(f\) does not depend on \(\Bl\), as the states are delocalized perpendicular to the electric field. We conjecture Eq.~\eqref{eqn:RMT_commens_ans} to be the statistical description of three-or-more-tone thermalizing quantum systems with continuous spectra in the incommensurate limit.

\section*{Acknowledgements}
	\label{sec:ack}

	The authors would like to thank V. Khemani, M. Kolodrubetz, and C. R. Laumann for helpful discussions. We also thank W. W. Ho, and D. Huse for comments on a draft of this article. Numerics were performed on the BU Shared Computing Cluster. DL and AC were supported by NSF Grant No. DMR-1752759, and AFOSR Grant No. FA9550-20-1-0235. PC's work at MIT was supported by the NSF STC ``Center for Integrated Quantum Materials'' under Cooperative Agreement No. DMR-1231319. This work was performed at the Aspen Center for Physics, which is supported by NSF Grant No. PHY-1607611.

\bibliography{qpavalanche}

\clearpage

\appendix
\section{Equivalence of Definitions of Quasiperiodically-Driven MBL}
	\label{app:equiv}

	There has already been a definition of quasiperiodically-driven MBL presented in the literature~\cite[Section II D]{Else2019}. The definition proposed in \autoref{sec:QPDMBL} is equivalent to that in Ref.~\cite{Else2019}.

	Ref.~\cite{Else2019} defines quasiperiodically-driven MBL by first supposing a decomposition of the evolution operator of the form
	\begin{equation}
		U(t,0) = P(\Bt_t) e^{-i t H_F} P(\Bt_0)^\dagger,
	\end{equation}
	where \(P(\Bt)\) is a quasilocal unitary. This is equivalent to our requirement of the existence of a complete set of smooth quasienergy states, as may be seen by taking
	\begin{equation}
		P(\Bt) = \sum_\alpha \ketbra{\phi_\alpha(\Bt)}{\alpha},\quad 
		H_F = \sum_\alpha \epsilon_\alpha \ketbra{\alpha}{\alpha},
	\end{equation}
	where some locality structure must be imposed on the basis \(\ket{\alpha}\) to make sense of \(P\) being quasilocal. For instance, the basis could be taken to be the product basis of uncoupled spins.

	Given this decomposition exists, Ref.~\cite{Else2019} defines a quasiperiodically-driven system to be MBL if there is a complete set of quasilocal integrals of motion for \(H_F\), which we can express in terms of the basis \(\ket{\alpha}\) as
	\begin{equation}
		\tau^{z\prime}_j = \sum_{\alpha} \tau^z_{j \alpha} \ketbra{\alpha}{\alpha}.
	\end{equation}
	The relation between this \(\tau^{z\prime}_j\) and our \(\tau^z_j(\Bt)\) is given by what Ref.~\cite{Else2019} calls ``reverse [Heisenberg] evolution''.
	\begin{equation}
		\tau^z_j(\Bt_t) = P(\Bt_t) \tau^{z\prime}_j P(\Bt_t)^\dagger
	\end{equation}
	Thus, the quasilocality of one of these objects implies the quasilocality of the other, and the two definitions of quasiperiodically-driven MBL are equivalent.

\section{Numerical Evidence For Thermal Ansatz}
	\label{app:numerical_ans}

	In this appendix we verify that the ansatz \eqref{eqn:RMT_ans} is effective for our purposes by numerically computing the matrix elements \(\tilde{V}_{\alpha \beta}^{\Bl}\) for \(\alpha \neq \beta\) and \(D = 2\) and checking that they obey the statistics we predict in Eq.~\eqref{eqn:RMT_ans}.

	\subsection{Model}
		\label{subsec:model_ans}

		We first define a model that we work with numerically. In principle, this should be a non-integrable many-body quantum system driven quasiperiodically. However, it has already been established numerically that the expectation values of operators in eigenstates of \emph{static} thermalizing Hamiltonians are well-described by random matrix theory, through ETH~\cite{Jensen1985,Deutsch,Srednicki1994,Rigol2008,DAlessio2016}. The content of our ansatz that requires new analysis is the frequency lattice structure.

		To separate the frequency lattice structure from a test of ETH, we choose a model that already consists of random matrices, and add quasiperiodic driving. The result is a Gaussian unitary ensemble (GUE) random Hamiltonian with nearest-neighbor hops on the frequency lattice. That is,
		\begin{equation}
			H(\Bt) = H_0 + J(H_1 e^{-i \theta_1} + H_2 e^{-i \theta_2} + \mathrm{h.c.}),
			\label{eqn:RMT_model}
		\end{equation}
		where \(H_0\) is a GUE random matrix with root-mean-square (rms) energy
		\begin{equation}
			\sqrt{\frac{1}{N}\tr{H_0^\dagger H_0}} = W + o(1),
			\label{eqn:rms_en}
		\end{equation}
		\(J\) sets the driving amplitude (and is a hopping amplitude in the frequency lattice), and \(H_1\) and \(H_2\) are complex Gaussian random matrices with unit rms energy. We take \(\theta_j = \Omega_j t\), with \(\Omega_1/\Omega_2 = (1+\sqrt{5})/2\) given by the golden ratio.

		We restrict our attention to the case of \(D=2\) tones, which is the most numerically tractable. As the ansatz~\eqref{eqn:RMT_ans} assumes no structure beyond that imposed by the assumption of localization and normalization, we expect that if the RMT phenomenology holds for \(D=2\) it will also hold for more tones, provided the larger \(D\) models are localized in the frequency lattice.

		We take \(V\) in Eq.~\eqref{eqn:RMT_ans} to be a static GUE random operator with unit rms energy.

	\subsection{Statistics of Matrix Elements in Commensurate Approximations}
		\label{subsec:stats_mat_els}

		Numerically extracting the quasienergy states \(\ket{\phi_\alpha(\Bt)}\) from the quasiperiodically driven model \eqref{eqn:RMT_model} can be challenging. It usually requires solving the model on the frequency lattice, which increases the size of the problem substantially. It is much easier instead to make a commensurate approximation to the incommensurate frequency vector \(\BO\) and solve the corresponding Floquet problem in the time domain. If the incommensurate model is localized in the frequency lattice, which is a requirement of our ansatz, then the incommensurate limit may be safely described by a limit of commensurate approximations~\footnote{Indeed, a commensurate approximation to \(\BO\) may be regarded as introducing periodic boundary conditions in the frequency lattice~\cite{Martin2017,Crowley2019,Long2021}.}.

		We consider commensurate approximations
		\begin{equation}
			\BO_n = \Omega_1 \hat{e}_1 + \Omega_1 \frac{p_n}{q_n} \hat{e}_2
		\end{equation}
		where \(p_n = F_{n-1}\) and \(q_n = F_n\) are consecutive Fibonacci numbers. As \(n \to \infty\), we have that \(\BO_n \to \BO\).

		Each commensurate approximation is periodic with period \(T_n = q_n \tfrac{2\pi}{\Omega_1}\). Thus, we can find the quasienergy states at \(\Bt=0\) and their corresponding quasienergies by diagonalizing the Floquet operator
		\begin{align}
			U(T_n,0) &= \mathcal{T}\exp\left(-i \int_0^{T_n} \d t\, H(\Bt_t) \right) \nonumber \\
			&= \sum_{\alpha} e^{-i\epsilon_\alpha T_n} \ketbra{\phi_\alpha(0)}{\phi_\alpha(0)},
		\end{align}
		where \(\mathcal{T}\) denotes time ordering. The quasienergy states at any other \(\Bt_t\) can then be calculated as
		\begin{equation}
			\ket{\phi_\alpha(\Bt_t)} = e^{i\epsilon_\alpha t} U(t,0) \ket{\phi_\alpha(0)}.
		\end{equation}

		\begin{figure}
			\includegraphics[width=\linewidth]{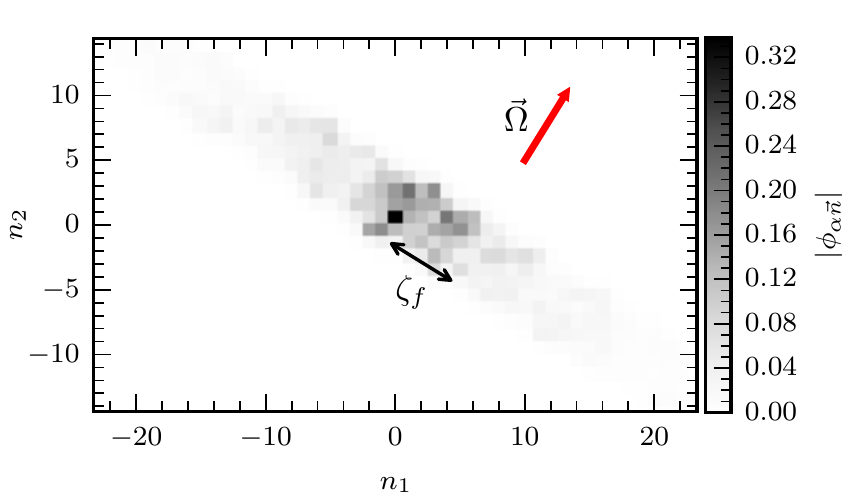}
			\caption{\label{fig:qst_frqlat}\emph{Frequency lattice quasienergy state.}--- Quasienergy states are well described by a sum \(\ket{\tilde{\phi}_\alpha}=\sum_{\Bn} \ket{\phi_{\alpha \Bn}}\ket{\Bn}\), where \(\ket{\phi_{\alpha \Bn}}\) are random vectors. The norm \(|\phi_{\alpha \Bn}|\) of the components decreases exponentially in the direction perpendicular to \(\BO\), with localization length \(\zeta_f\). They decrease faster than exponentially parallel to \(\BO\). \emph{Parameters:} \(N=20\), \(J/W = 0.1\), \(\Omega_1/W = 0.6\), \(q = 233\)}
		\end{figure}

		We use a second-order Suzuki-Trotter approximation~\cite{Wiebe2010} to compute \(U(T_n,0)\), and subsequently calculate \(\ket{\phi_\alpha(\Bt)}\) on an \(F_{n-1} \times F_{n}\) grid in the \(\Bt\) torus. We fix a gauge for this state by requiring that the highest weight component in the corresponding frequency lattice state \(\ket{\tilde{\phi}_\alpha} = \sum_\Bn \ket{\phi_{\alpha\Bn}} \ket{\Bn}\) be \(\ket{\phi_{\alpha 0}}\). In this gauge, if the quasienergy states are well-localized, we may regard our chosen representative states as being centered at the origin in the frequency lattice~(\autoref{fig:qst_frqlat}).

		With \(\ket{\phi_\alpha(\Bt)}\) found, we can compute the matrix elements \(\tilde{V}_{\alpha \beta}^{\Bl}\) as the two-dimensional Fourier coefficients of
		\begin{equation}
			V_{\alpha \beta}(\Bt) = \bra{\phi_\alpha(\Bt)}V(\Bt)\ket{\phi_\beta(\Bt)}.
		\end{equation}

\begin{figure}
			\centering
			\includegraphics[width=\linewidth]{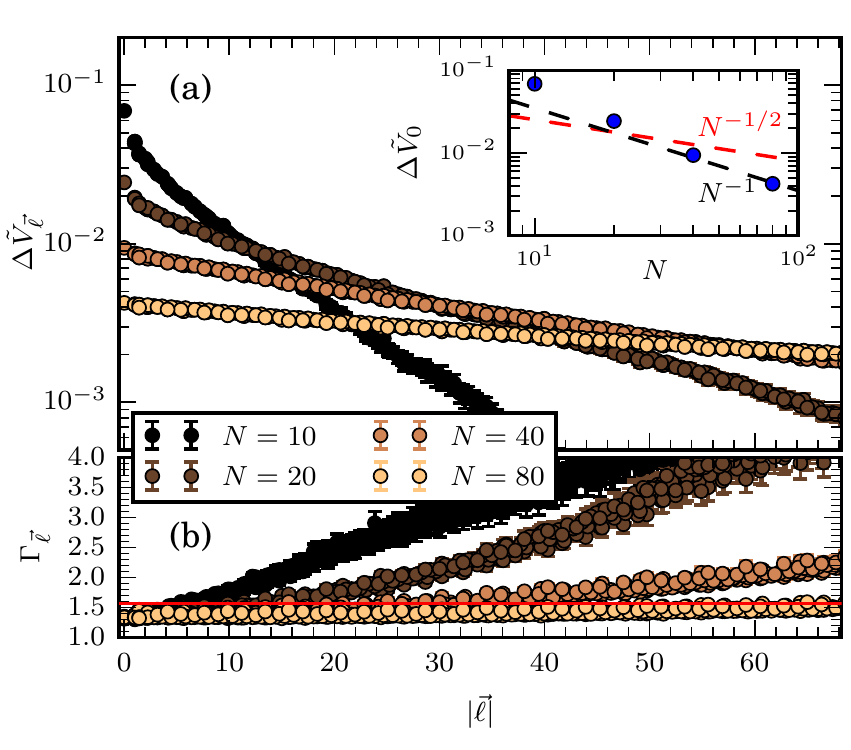}
			\caption{\label{fig:ansatz_perp}\emph{Matrix elements perpendicular to \(\BO\).}--- We examine the statistics of matrix elements \(\tilde{V}_{\alpha \beta}^{\Bl}\) for \(|\Bl\cdot \BO_n| < W\) (almost perpendicular to the electric field). (\textbf{a})~Eq.~\eqref{eqn:RMT_ans} conjectures that the matrix elements should have a standard deviation which decreases exponentially in \(|\Bl|\). This feature is visible for small \(N\). (Inset) The standard deviation of the \(\Bl = 0\) matrix elements decreases faster than \(N^{-1/2}\) (red dashed), as we predict. \autoref{sec:synth} predicts a scaling of \(N^{-1}\) (black dashed), which is a better fit for large \(N\), though Appendix~\ref{app:numerics_synth_loc} provides much better evidence for this scaling. (\textbf{b})~The ratio \(\Gamma_{\Bl}\) should be \(\pi/2\) (red line) for a Gaussian distribution of matrix elements. We see this is not the case for large \(|\Bl|\). \emph{Parameters:} \(J/W = 0.1\), \(\Omega_1/W = 0.6\), \(q = 233\), with \(N_{\mathrm{samp}} \approx 1200/N\) random matrix samples.}
		\end{figure}

		Using this method, we can directly compute the matrix elements \(\tilde{V}_{\alpha \beta}^{\Bl}\) in small commensurate approximations. We address the behavior of the matrix elements for \(\Bl\) perpendicular to and parallel to the electric field \(\BO\) separately. We begin with \(\Bl \perp \BO\) (\autoref{fig:ansatz_perp}).

		We have assumed that the standard deviation \(\Delta \tilde{V}_\Bl\) of the matrix elements with fixed \(\Bl\) should decrease exponentially for large \(|\Bl|\) in this direction (the mean vanishes). Specifically, we predict for the off-diagonal matrix elements that
		\begin{equation}
		 	\Delta \tilde{V}_\Bl \sim \frac{\|f_\Bl\|}{\sqrt{N_{\mathrm{eff}}}},
		\end{equation} 
		where
		\begin{equation}
			\|f_\Bl\|^2 = \int\d\omega\, |f_\Bl(\omega)|^2
		\end{equation}
		decays exponentially.
		This exponential decay is visible for small \(N\) in \autoref{fig:ansatz_perp}(a), but we are unable to reach commensurate approximations that allow us to see the decay clearly for larger \(N\).

		We can also observe that \(\Delta \tilde{V}_0\) decays faster than \(N^{-1/2}\) for fixed \(\Bl=0\). This is also predicted by our ansatz, as the localization length \(\xi_f\) may grow with \(N\), so that \(N_{\mathrm{eff}}\) grows faster than \(N\). Indeed, \autoref{sec:synth} gives that \(\xi_f = O(N)\) for \(D=2\), so that \(\Delta \tilde{V}_0 = O(N^{-1})\).

		We did not require that the matrix elements be normally distributed, as is often done in ETH. Indeed, in the tails of a localized wavefunction the wavefunction amplitudes, and hence matrix elements, should be log-normally distributed~\cite{Scardicchio2017}. We can check if the matrix elements we compute numerically are normally distributed by computing the ratio~\cite{LeBlond2019}
		\begin{equation}
			\Gamma_\Bl = \frac{\cexp{|V_{\alpha\beta}^{\Bl}|^2}}{\cexp{|V_{\alpha\beta}^{\Bl}|}^2},
		\end{equation}
		where angle brackets indicate an average over off-diagonal elements \(V_{\alpha\beta}^{\Bl}\) for fixed \(\Bl\), and within a window of the quasienergy difference \(\Delta_{\alpha\beta}^{\Bl}\). \(\Gamma_\Bl\) is \(\pi/2\) if the matrix elements are Gaussian-distributed for fixed \(\Bl\), within a small quasienergy window.

		\begin{figure}
			\centering
			\includegraphics[width=\linewidth]{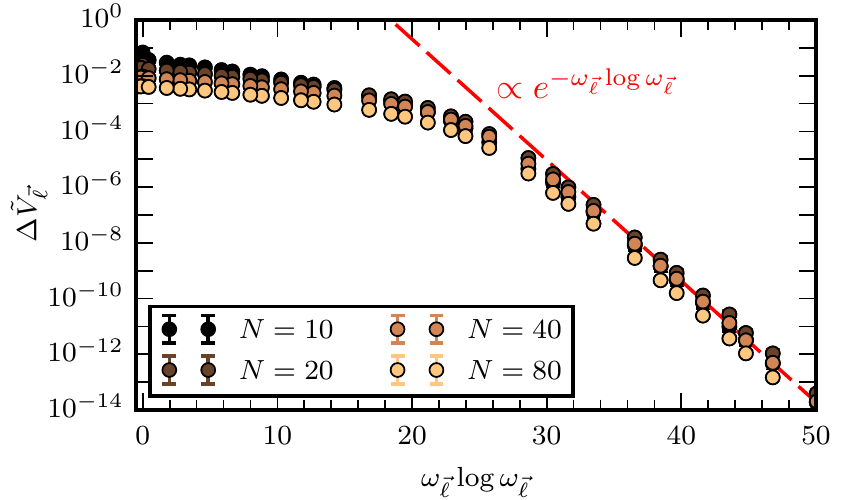}
			\caption{\label{fig:ansatz_par}\emph{Matrix elements parallel to \(\BO\).}--- We examine the statistics of matrix elements \(\tilde{V}_{\alpha \beta}^{\Bl}\) for \(|\Bl\times \BO_n| < |\BO_n|\) (almost parallel to the electric field). Eq.~\eqref{eqn:nlogn} predicts that the standard deviation of the matrix elements should decay much faster as compared to the perpendicular direction. This is reflected in our numerics, where the decay is faster than exponential: consistent with \(\log \Delta \tilde{V}_\Bl \sim -\omega_\Bl \log \omega_\Bl\) (where \(\omega_\Bl = \Bl\cdot\hat{\Omega}\)) for large \(\omega_\Bl\). Furthermore, the localization length of the matrix elements does not grow with \(N\) -- increasing \(N\) only decreases the matrix elements. \emph{Parameters:} \(J/W = 0.1\), \(\Omega_1/W = 0.6\), \(q = 233\), with \(N_{\mathrm{samp}} \approx 1200/N\) random matrix samples.}
		\end{figure}

		We see in \autoref{fig:ansatz_perp}(b) that most matrix elements are not Gaussian-distributed. For small deviations from \(\pi/2\), this may be because the windows we have used for \(\Delta_{\alpha\beta}^{\Bl}\) are too large. Taking smaller windows while still maintaining good statistics requires larger \(N\). The large deviations visible at small \(N\) and large \(|\Bl|\) cannot be explained in this way; they represent departures from Gaussianity.

		Our ansatz predicts qualitatively different behavior of the matrix elements with \(|\Bl|\) when \(\Bl\) is parallel to the electric field \(\BO\). These predictions are verified in \autoref{fig:ansatz_par}. Namely, the standard deviation \(\Delta\tilde{V}_\Bl = O(e^{-\omega_\Bl \log \omega_\Bl})\) decreases faster than exponentially for \(\omega_\Bl \gg \xi_\parallel\), and \(\xi_\parallel\) does not depend on \(N\). (For a typical spin system with high-frequency quasiperiodic driving \(\xi_\parallel\) should depend weakly on \(N\), because the bandwidth of the static part of the Hamiltonian grows. Our model~\eqref{eqn:RMT_model} has a fixed bandwidth, so \(\xi_\parallel\) should not depend on \(N\).)

		Indeed, faster-than-exponential decay of \(\Delta\tilde{V}_\Bl\) is visible in \autoref{fig:ansatz_par}. Furthermore, increasing \(N\) only decreases \(\Delta\tilde{V}_\Bl\) (due to the factor \(N_{\mathrm{eff}}^{-1/2}\)), without extending the localization length \(\xi_\parallel\).

		The features of the ansatz listed in this appendix are those most relevant for this paper. We have verified that they are effective descriptions of the frequency lattice structure of matrix elements in the localized (in the synthetic dimensions) regime.

\section{Typical Frequency Lattice Fidelity Susceptibility}
	\label{app:chi_star}

	In this appendix, we prove Eq.~\eqref{eqn:chi_star}. Restated here, we show that the fidelity susceptibilities of frequency lattice eigenstates are distributed according to a power law
	\begin{equation}
		f_{\mathrm{FS}}(\chi) \sim \sqrt{\frac{\chi_{\star,j}}{\chi^3}}
	\end{equation}
	where the typical scale is
	\begin{equation}
		\sqrt{\chi_{\star,j}} = \lim_{\Delta \to 0}\left[\frac{1}{2\Delta}\sum_{|\Delta_{\beta\alpha}^{\Bl}|<\Delta} |\tilde{V}^{\Bl}_{\beta\alpha}|\right].
		\label{eqn:app_chi_star}
	\end{equation}

	We split the sum in \eqref{eqn:fid_sup} into a sum for each frequency lattice site, \(\chi_\alpha = \sum_{\Bl} \chi_{\alpha,\Bl},\) where
	\begin{equation}
		\chi_{\alpha,\Bl} = \sum_{\beta,h} \left|\frac{\tilde{V}_{\beta\alpha}^{\Bl}}{\omega_{\beta\alpha} + \Bl\cdot\BO - 2h}\right|^2.
	\end{equation}
	Due to the presence of small denominators \(\Delta_{\beta \alpha}^{\Bl} = |\omega_{\beta\alpha} + \Bl\cdot\BO - 2h|\), this sum tends to be dominated by its largest element. Then we can write
	\begin{equation}
		\chi_{\alpha,\Bl} \approx \frac{|\tilde{V}_{\beta\alpha}^{\Bl}|^2}{|\Delta_{\beta \alpha}^{\Bl}|^2},
	\end{equation}
	where \(\beta\) and \(h\) are chosen to minimize \(|\Delta_{\beta \alpha}^{\Bl}|^2\). The distribution of fidelity susceptibilities \(f_{\mathrm{FS}}(\chi|\omega_{\beta\alpha},\Bl)\) can then be calculated as~\cite{Crowley2021}
	\begin{align}
		 f_{\mathrm{FS}} &= \int\d V \int\d \Delta \, \delta\left(\chi-\tfrac{|V|^2}{|\Delta|^2}\right) f_{\mathrm{ME}}(V) f_{\mathrm{LS}}(\Delta) \\
		 &= \frac{1}{2\chi^{3/2}}\int\d V \,|V| f_{\mathrm{ME}}(V) f_{\mathrm{LS}}(\tfrac{V}{\sqrt{\chi}})
	\end{align} 
	where \(f_{\mathrm{ME}}\) and \(f_{\mathrm{LS}}\) are distributions for the matrix element and minimum level spacing \(\Delta_{\beta \alpha}^{\Bl}\) respectively. Both depend on \(\Bl\). This calculation shows that \(f_{\mathrm{FS}} \sim \sqrt{\chi_{\star,\Bl}/\chi^3}\) has a power-law dependence on \(\chi\). The scale \(\chi_{\star,\Bl}\) may be extracted as
	\begin{align}
		\chi_{\star,\Bl} &= \lim_{\chi\to\infty} \chi^3 f_{\mathrm{FS}}^2 \\
		&= \left(\lim_{\Delta\to 0}\frac{1}{2}\int\d V\, |V| f_{\mathrm{ME}}(V) f_{\mathrm{LS}}(\Delta)\right)^2.
	\end{align}
	Schematically, this may be written \(\chi_{\star,\Bl} = [|V^\Bl|]^2 \rho_\Bl^2\), where \([|V^\Bl|]\) is an average of the absolute value of the matrix elements \(\tilde{V}_{\beta\alpha}^{\Bl}\) as the random variables \(R_{\beta \alpha,\Bl}\) from Eq.~\eqref{eqn:RMT_ans} are varied. The quantity \(\rho_\Bl\) is a density of states at the relevant quasienergy, restricted to the site \(\Bl\). However, it will be more useful later to instead express \(\sqrt{\chi_{\star,\Bl}}\) explicitly as
	\begin{equation}
		\sqrt{\chi_{\star,\Bl}} = \lim_{\Delta \to 0}\left[\frac{1}{2\Delta}\sum_{|\Delta_{\beta\alpha}^{\Bl}|<\Delta} |\tilde{V}^{\Bl}_{\beta\alpha}|\right],
	\end{equation}
	where \(\Bl\) is fixed in the sum, and square brackets indicate an average over the variables \(R_{\beta \alpha,\Bl}\) and over the quasienergies \(\epsilon_\beta-2h\). We have not specified the distributions \(f_{\mathrm{ME}}\) and \(f_{\mathrm{LS}}\) over which this average is to be performed because for our purposes all we require is that the average \([|V^\Bl|]\) exists, and that the probability density \(f_{\mathrm{LS}}(0)\) is finite. The specific distribution of the matrix elements and quasienergies will affect the value of \(\sqrt{\chi_{\star,\Bl}}\), but not its asymptotic scaling as the avalanche progresses, which is our only concern.

	If we then make the approximation that the random variables \(\chi_{\alpha,\Bl}\) on different sites are independent, we can calculate the typical scale \(\chi_{\star,j}\) of \(\chi_\alpha\) in terms of the distributions on the sites \(\Bl\). We define the cumulant generating functions
	\begin{equation}
		K_\Bl(t) = \log [e^{i\chi_{\alpha,\Bl} t}],
	\end{equation}
	where the square brackets indicate an average over \(\chi_{\alpha,\Bl}\), appropriately weighted by the distribution \(f_{\mathrm{FS}}\).

	As the asymptotic form of the fidelity distribution is \(f_{\mathrm{FS}} \sim \chi_{\star,\Bl}^{1/2}/\chi^{3/2}\) for \(\chi\to\infty\), the cumulant generating function must behave asymptotically for \(t \to 0\) as
	\begin{equation}
		K_\Bl(t) \sim C\sqrt{t \chi_{\star,\Bl}},
	\end{equation}
	where \(C = (-1+i)\sqrt{2\pi}\) is a constant~\cite{Crowley2021}. The cumulant generating function for a sum of independent random variables is the sum of their cumulant generating functions, thus
	\begin{equation}
		K(t) \sim C\sqrt{t} \sum_{\Bl} \sqrt{\chi_{\star,\Bl}}.
	\end{equation}
	That is, we have a full distribution of \(\chi_\alpha\) with the same power-law tail, and a scale \(\chi_{\star,j}\) given by
	\begin{equation}
		\sqrt{\chi_{\star,j}} = \sum_{\Bl} \sqrt{\chi_{\star,\Bl}} = \lim_{\Delta \to 0}\left[\frac{1}{2\Delta}\sum_{|\Delta_{\beta\alpha}^{\Bl}|<\Delta} |\tilde{V}^{\Bl}_{\beta\alpha}|\right].
	\end{equation}
	The sum is over \(\ket{\tilde{\phi}_\beta^\Bl \{\tilde{\tau}'\}}\) satisfying the condition \(|\Delta_{\beta\alpha}^{\Bl}|<\Delta\). With fixed \(\Bl\) this sum is finite, with at most \(2N\) terms for any \(\Bl\). The infinite sum over \(\Bl\) converges if \(|\tilde{V}^{\Bl}_{\beta\alpha}|\) decays exponentially in \(|\Bl|\), as we have assumed. The \(\Delta \to 0\) limit converges if \(f_{\mathrm{LS}}(0)\) is finite for all \(\Bl\). Thus, \(\sqrt{\chi_{\star,j}}\) is a finite quantity for any \(N, \xi_f < \infty\).

	Let us return to the assumption that the random variables \(\chi_{\alpha,\Bl}\) are independent for different \(\Bl\). The matrix elements \(\tilde{V}_{\beta \alpha}^{\Bl}\) appearing at distinct \(\Bl\) are independent random variables within our ansatz, but the energy denominators \(\Delta_{\beta \alpha}^{\Bl}\) do have correlations between them. These correlations arise because the change in a given energy denominator is given deterministically by the change in the \(\Bl \cdot \BO\) term. This results in special separations \(\Bl_*\) where \(\Bl_*\cdot \BO \approx 0\), and so the energy denominators are almost the same. On this point, we observe that these special \(\Bl_*\) occur no more frequently than would be expected for random shifts in quasienergy, so even if they do introduce some correlation, it is unlikely to affect the asymptotic behavior we have identified.

	In more detail, for a badly approximable \(\BO \in \reals^2\), there is a \(C>0\) such that~\cite{Schmidt1972,Schmidt1996,Else2019}
	\begin{equation}
		|\Bl \cdot \BO| \geq \frac{C |\BO|}{|\Bl|}.
	\end{equation}
	(A similar statement may be made for almost all \(\BO \in \reals^2\) by replacing \(|\Bl|\) with \(|\Bl|^{1+\epsilon}\) for any \(\epsilon >0\).) Thus, if \(|\Bl_* \cdot \BO| < \delta\) is especially small, then
	\begin{equation}
		|\Bl_*| \geq C|\BO|/\delta = O(\delta/|\BO|)^{-1}.
	\end{equation}
	In words, to find a potential \(\Bl_*\cdot\BO\) that is smaller than \(\delta\), one must search within a distance \(O(\delta/|\BO|)^{-1}\) in the frequency lattice. Similarly, if the potentials \(\Bl\cdot \BO\) were actually random, one would expect to have to sample \(O(\delta/|\BO|)^{-1}\) of them to find one that is smaller than \(\delta\).

\section{Numerical Evidence of Synthetic Localization for Two-Tone Driving}
	\label{app:numerics_synth_loc}

	Our calculations in \autoref{sec:synth} on the behavior of \(\xi_f\) with \(N\) for \(D=2\) can be verified through a number of numerical experiments. In this appendix, we report on two such experiments, one based on real-time evolution in a sequence of commensurate approximations to the quasiperiodic drive (Appendix~\ref{subsec:comens}), and one based on the one-dimensional model \eqref{eqn:K1dim} in the frequency lattice (Appendix~\ref{subsec:quasi1d}). In both cases, our results are consistent with \(\xi_f = O(\zeta_f) = O(N)\).

	\subsection{Model}
		\label{subsec:model_synth}

		We use the model \eqref{eqn:RMT_model} from Appendix~\ref{subsec:model_ans} for our numerics. This is a model of driven random matrices with nearest-neighbor hops on the frequency lattice. Restating it here:
		\begin{equation}
			H(\Bt) = H_0 + J(H_1 e^{-i \theta_1} + H_2 e^{-i \theta_2} + \mathrm{h.c.}),
			\label{eqn:RMT_model2}
		\end{equation}
		where \(H_0\) is a GUE random matrix with rms energy \(W\) (as defined in Eq.~\eqref{eqn:rms_en}), \(J\) is a hopping amplitude, and \(H_1\) and \(H_2\) are complex Gaussian random matrices with unit rms energy. We take \(\theta_j = \Omega_j t\), with \(\Omega_1/\Omega_2 = (1+\sqrt{5})/2\) given by the golden ratio.

	\subsection{Commensurate Approximations}
		\label{subsec:comens}

		Ideally, we could directly compute the quasienergy states \(\ket{\phi_\alpha(\Bt)}\) from the quasiperiodically driven model \eqref{eqn:RMT_model}, but as we noted in Appendix~\ref{subsec:stats_mat_els}, this is numerically challenging, and so instead we make a commensurate approximation to the incommensurate frequency vector \(\BO\), and we solve the corresponding Floquet problem.

		Recall that the commensurate approximations we use are
		\begin{equation}
			\BO_n = \Omega_1 \hat{e}_1 + \Omega_1 \frac{p_n}{q_n} \hat{e}_2
		\end{equation}
		where \(p_n = F_{n-1}\) and \(q_n = F_n\) are consecutive Fibonacci numbers. We use a second-order Suzuki-Trotter approximation~\cite{Wiebe2010} to compute \(U(T_n,0)\), and subsequently calculate \(\ket{\phi_\alpha(\theta_2 \hat{e}_2)}\) at \(q\) points along the line \(\theta_1 = 0\).

		Localization in the frequency lattice can be probed by calculating the Fourier coefficients of the density matrix
		\begin{equation}
			\rho_{\alpha}(\theta_2 \hat{e}_2) = \ketbra{\phi_\alpha(\theta_2 \hat{e}_2)}{\phi_\alpha(\theta_2 \hat{e}_2)} = \sum_{n} \rho'_{\alpha n} e^{-i n \theta_2},
		\end{equation}
		which are related to the two-dimensional Fourier coefficients of the density matrix \(\rho_{\alpha \Bn}\) by
		\begin{equation}
			\rho'_{\alpha n} = \sum_{n_1} \rho_{\alpha,n_1\hat{e}_1+n\hat{e}_2}.
		\end{equation}
		Computing \(\rho'_{\alpha n}\), rather than \(\rho_{\alpha \Bn}\), is less expensive numerically (in both time and memory), and allows us to probe larger commensurate approximations. We calculate the density matrix, rather than the kets \(\ket{\phi_\alpha(\Bt)}\), to avoid having to find a smooth gauge for the states.

		To quantify the localization of these states, we use the inverse participation ratio, defined as
		\begin{equation}
			\mathrm{IPR}_\alpha = \sum_{n} \|\rho'_{\alpha n}\|_F^4,
			\label{eqn:IPR}
		\end{equation}
		where \(\|\cdot\|_F\) is the Frobenius norm. This quantity is \(1\) for a perfectly localized state, and \(1/q\) for a completely delocalized state on \(q\) sites. (We do not have an infinite system as we calculate \(\rho_{\alpha}(\theta_2 \hat{e}_2)\) at only \(q\) points.) Roughly, \(1/\mathrm{IPR}_\alpha\) is the number of frequency lattice sites that a state has significant weight on, and is proportional to \(\zeta_f\), the localization length of the quasienergy states. As we observed in \autoref{sec:ansatz}, the localization length of the matrix elements has the same scaling: \(\xi_f = O(\zeta_f)\). Thus, it is sufficient to compute \(\zeta_f\).

		The numerically calculated inverse participation ratios for the model \eqref{eqn:RMT_model} are shown in \autoref{fig:commen_IPR}. For every \(N\) in \autoref{fig:commen_IPR}, the \(\mathrm{IPR}\) saturates as \(q\) becomes very large, indicating that all \(N\) have a finite localization length, as we have predicted.

		Furthermore, rescaling \(q\) by \(1/N\) and the \(\mathrm{IPR}\) by \(N\) produces a collapse of the data. This amounts to rescaling lengths in the frequency lattice by \(1/N\), so the data collapse indicates the existence of a single length scale, \(\xi_f = O(\zeta_f)\), which grows proportionally to \(N\). Thus, these numerics agree with our prediction of \(\nu = 1\).

	\subsection{One-dimensional Approximation}
		\label{subsec:quasi1d}

		We can probe even larger distances in the frequency lattice, and larger Hilbert space dimensions \(N\), by instead studying the one-dimensional approximation \eqref{eqn:K1dim} directly.

		There are many numerical methods effective in solving one-dimensional tight binding models. For the purpose of extracting the localization length \(\zeta_f\) (which has the same scaling as \(\xi_f\)), we use a transfer matrix method~\cite{Matsuda1970,Thouless1974}.

		The eigenvalue equation for \(\ket{\tilde{\phi}_\alpha} = \sum_{k} \ket{\phi_{\alpha,k}}\ket{k}\) may be written
		\begin{equation}
			(H_0 + \omega_k) \ket{\phi_{\alpha,k}} + J(H_{i_k} \ket{\phi_{\alpha,k-1}} + H_{i_{k+1}}^\dagger\ket{\phi_{\alpha,k+1}}) = \epsilon_\alpha \ket{\phi_{\alpha,k}},
			\label{eqn:eig_components}
		\end{equation}
		where \(i_k \in \{1,2\}\) is the same quasiperiodic sequence from \autoref{sec:synth}, and \(H_0\), \(H_1\) and \(H_2\) are given as in the model~\eqref{eqn:RMT_model}.

		The eigenvalue equation~\eqref{eqn:eig_components} may be expressed as a transfer matrix equation for \(\ket{\phi_{\alpha,k+1}}\) given \(\ket{\phi_{\alpha,k}}\) and \(\ket{\phi_{\alpha,k-1}}\):
		\begin{equation}
			\ket{\Phi_{\alpha,k+1}} = \begin{pmatrix}
				\ket{\phi_{\alpha,k+1}} \\ \ket{\phi_{\alpha,k}}
			\end{pmatrix}
			= 
			T_{k+1}(\epsilon_\alpha)
			\begin{pmatrix}
				\ket{\phi_{\alpha,k}} \\ \ket{\phi_{\alpha,k-1}}
			\end{pmatrix},
		\end{equation}
		where
		\begin{equation}
			T_{k+1}(\epsilon) = 
			\begin{pmatrix}
				-\tfrac{1}{J}H^{-\dagger}_{i_{k+1}}(H_0 + \omega_k - \epsilon) & -H^{-\dagger}_{i_{k+1}} H_{i_k} \\
				\mathbbm{1} & 0
			\end{pmatrix},
		\end{equation}
		and we have written \(A^{-\dagger} = (A^{-1})^\dagger = (A^\dagger)^{-1}\).

		\begin{figure}
			\centering
			\includegraphics[width=\linewidth]{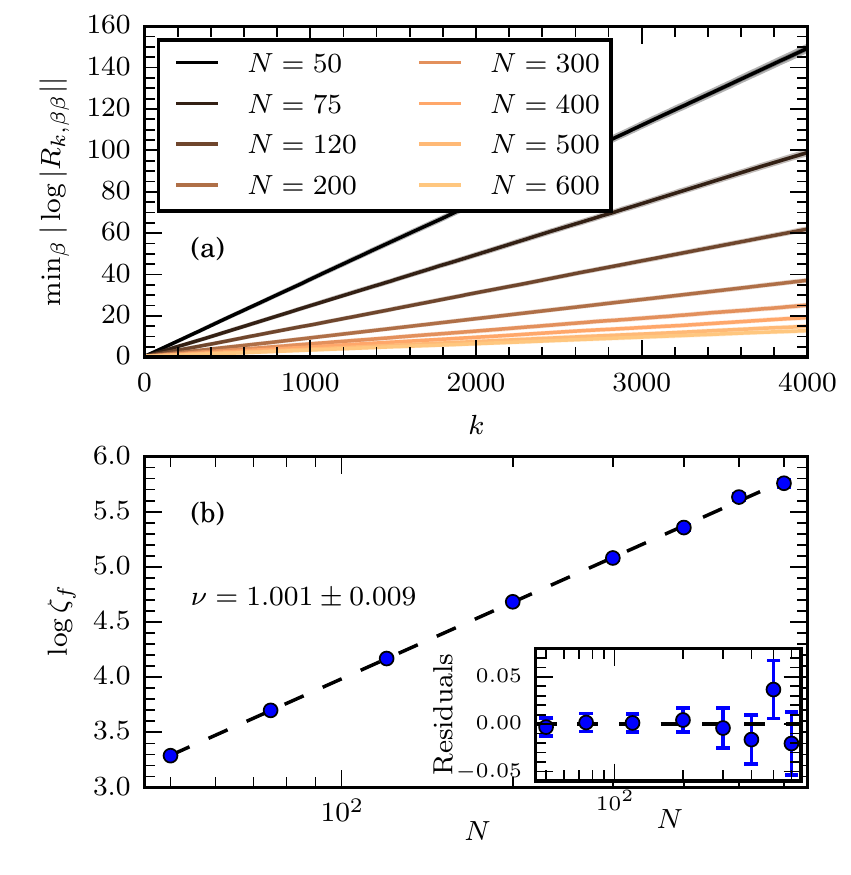}
			\caption{\label{fig:shooting}\emph{Scaling of \(\zeta_f\) in a one-dimensional approximation.}--- (\textbf{a})~The inverse localization length \(\zeta_f^{-1}\) may be extracted from the limiting behavior of the QR-decomposition of a transfer matrix, as described in~\eqref{eqn:shoot_good}. (\textbf{b})~Values of \(\zeta_f\) extracted from the data in (\textbf{a}) show the expected linear scaling with \(N\). Fitting a power law \(\zeta_f = A N^\nu\) (dashed line) gives \(\nu = 1.001\pm 0.009\), consistent with \(\nu=1\). \emph{Parameters:} \(\Omega_1/W=0.6\), \(J/W = 0.2\), \(\Omega_1/\Omega_2 = (1+\sqrt{5})/2\), \(\epsilon = 0\), chain length \(L = 4000\) with between \(800\) and \(200\) samples of random matrices, depending on \(N\).}
		\end{figure}

		To identify the localization length \(\zeta_f\), we need to identify the asymptotic behavior
		\begin{equation}
			\zeta_f^{-1} = \lim_{k \to \infty} -\frac{1}{k}\log\|\Phi_{\alpha, k}\|.
		\end{equation}
		The scaling of \(\log\|\Phi_{\alpha, k}\|\) can be estimated by computing the eigenvalues of 
		\begin{equation}
			\Pi_k(\epsilon) = T_k T_{k-1} \cdots T_1(\epsilon)
		\end{equation}
		at a fixed target quasienergy \(\epsilon\). The product \(\Pi_k\) has \(2N\) eigenvalues \(\lambda_{k\beta}\), which may have \(|\lambda_{k\beta}|<1\) corresponding to decay of the wavefunction, or \(|\lambda_{k\beta}|>1\) corresponding to growth of the wavefunction (moving towards the localization center). The longest localization length is extracted as
		\begin{equation}
			\zeta_f^{-1}(\epsilon) = \lim_{k \to \infty} \min_\beta \frac{1}{k}\left|\log|\lambda_{k\beta}|\right|.
			\label{eqn:shoot_bad}
		\end{equation}

		Equation \eqref{eqn:shoot_bad} is hard to evaluate numerically, as the eigenvalues of \(\Pi_k(\epsilon)\) vary over many orders of magnitude for large \(k\), and numerical calculations tend to be dominated by the largest eigenvalue. Fortunately, numerically stable methods to calculate \(\zeta^{-1}_f(\epsilon)\) have been developed in the context of calculating Lyapunov exponents in discrete maps~\cite{Geist1990}. They are based on the QR-decomposition of \(\Pi_k\),
		\begin{equation}
			\Pi_k = Q_k R_k
		\end{equation}
		where \(Q_k\) is unitary and \(R_k\) is upper triangular. The localization length may be computed as
		\begin{equation}
			\zeta_f^{-1}(\epsilon) = \lim_{k \to \infty} \min_\beta \frac{1}{k}\left|\log|R_{k,\beta\beta}|\right|,
			\label{eqn:shoot_good}
		\end{equation}
		where \(R_{k,\beta\beta}\) is a diagonal element of \(R_k\). By using the techniques of~\cite{Geist1990}, the logarithms \(\log|R_{k,\beta\beta}|\) may be computed directly. These are not dominated by the largest value, as the exponential growth with \(k\) in the elements \(R_{k,\beta\beta}\) appears only as linear growth in the logarithm.

		Localization lengths extracted using the transfer matrix method for different values of \(N\) and a value of \(\epsilon\) in the middle of the spectrum are shown in \autoref{fig:shooting}. (There is no ``middle of the spectrum'' in the full frequency lattice model, where the spectrum is unbounded. However, when restricted to a line as is in this section, the spectrum is bounded, and so it has a ``middle'' where the density of states is maximal, and the one-dimensional model is most representative of the frequency lattice.) We first see that the average of \(\min_\beta \left|\log|R_{k,\beta\beta}|\right|\) over random matrix samples (and even the individual samples, not shown) shows linear behavior with \(k\) with a strictly positive slope, so there is indeed exponential localization. Extracting the localization length from these data and fitting a power law \(\zeta_f = A N^\nu\) gives
		\begin{equation}
			\nu = 1.001 \pm 0.009,
		\end{equation}
		consistent with the predicted \(\nu=1\) from the associated Anderson model.

\end{document}